# Spin Parameters and Shape Models of Near-Earth Asteroids (4660) Nereus, (21088) Chelyabinsk, (66146) 1998 TU3, and (297418) 2000 SP43

Javier Rodríguez Rodríguez[1,2] , Enrique Díez Alonso[1,2] , Santiago Iglesias Álvarez[1] , Saúl Pérez Fernández[1,3] ,
Alejandro Buendia Roca[1,2] , Julia Fernández Díaz[1] , Javier Licandro[4,5] , Miguel R. Alarcon[4,5] , Miquel Serra-Ricart[4,5,6] ,
Amadeo Aznar Macías[7] , and Francisco Javier de Cos Juez[1,8]

[1] Instituto Universitario de Ciencias y Tecnologías Espaciales de Asturias (ICTEA), University of Oviedo, C. Independencia 13, 33004 Oviedo, Spain;
rodriguezrjavier@uniovi.es
[2] Departamento de Matemáticas, Facultad de Ciencias, Universidad de Oviedo, 33007 Oviedo, Spain
[3] Departamento de Informática, Facultad de Ciencias, Universidad de Oviedo, 33007 Oviedo, Spain
[4] Instituto de Astrofísica de Canarias (IAC), C/Vía Láctea sn, 38205, La Laguna, Spain; jlicandr@iac.es
[5] Departamento de Astrofísica, Universidad de La Laguna, 38206, La Laguna, Tenerife, Spain
[6] Light Bridges S. L., Observatorio Astronómico del Teide. Carretera del Observatorio del Teide, s/n, Güímar, Santa Cruz de Tenerife, Canarias, Spain
[7] Observatorio Isaac Aznar-MPC Z95, Alcublas, Valencia, Spain
[8] Departamento de Explotación y Prospección de Minas, Universidad de Oviedo, 33004 Oviedo, Spain
Received 2025 July 2; revised 2025 November 29; accepted 2025 December 4; published 2026 January 21

## Abstract

We present our shape and rotational characterization of four near-Earth asteroids observed as part of the Visible NEAs Observations Survey. This work includes 61 new light curves obtained between 2020 June and 2024 September for asteroids (21088) Chelyabinsk and (66146) 1998 TU3 and the potentially hazardous asteroids (4660) Nereus and (297418) 2000 SP43. We combine these observations with archival data to derive their shape models, spin parameters, and refined rotation periods using a light-curve inversion method. Our derived rotational periods are consistent with previously published values for all the objects. We present constant-period models for each asteroid, while for (66146) 1998 TU3 we detect a possible Yarkovsky–O'Keefe–Radzievskii–Paddack (YORP) acceleration of $\upsilon \simeq 2.05 \times 10^{-8}$ rad day$^{-2}$, potentially adding it to the short list of asteroids known to experience the YORP effect.

*Unified Astronomy Thesaurus concepts:* Asteroids (72); Near-Earth objects (1092); Photometry (1234); Apollo group (58); Amor group (36); Aten group (110)

## 1. Introduction

In this work, we focus our efforts on characterizing near-Earth asteroids (NEAs). According to the Center for Near-Earth Object Studies (CNEOS[9]), NEAs are those asteroids whose orbits have perihelion distances (q) <1.3 au. NEAs are classified into four groups: Atira, Atens, Apollo, and Amor, depending on their semimajor axes (a) and perihelion distances (q). As of 2025 February 24, there are 37,698 NEAs according to CNEOS,[10] a number that keeps actually increasing at a rate of ~3000 new NEAs each year. NEAs are also classified by their size and their threat of collision with Earth; an NEA is considered a potentially hazardous asteroid (PHA) if its Earth minimum orbit intersection distance (MOID) is ⩽0.05 au and its absolute magnitude (H) is ⩽22 (which implies a diameter D > 140 m).

The increasing number of discoveries[11] of NEAs and PHAs is due to several sky patrolling surveys, such as the past Lincoln Near-Earth Asteroid Research (LINEAR; G. H. Stokes et al. 2000), the Wide-field Infrared Survey Explorer (WISE; E. L. Wright & WISE Team 2009), or the current Panoramic Survey Telescope and Rapid Response System (Pan-STARRS; K. C. Chambers et al. 2018), Asteroid Terrestrial-impact Last Alert System (ATLAS; J. L. Tonry et al. 2018), and Near-Earth Object WISE (NEOWISE; A. Mainzer et al. 2011a).

To achieve our goal of asteroid shape modeling and spin parameter characterization, we applied the convex inversion method (M. Kaasalainen & J. Torppa 2001; M. Kaasalainen et al. 2001). With this method, from a suitable set of light curves, a convex shape model and the asteroid's spin parameters can be computed. We understand "as a suitable set of light curves" a number of observations acquired at different epochs, with varying solar phase angle $\alpha$ and aspect angle $\theta$ (the angle between the observer's line of sight and the asteroid's spin axis). The process we follow to achieve this is explained in Section 3, while the new photometric observations are presented in Section 2.

The convex inversion method can work with two types of light curves, dense and sparse. A dense light curve is that observed in a row, and with a high cadence, on the order of minutes like the data from Visible NEAs Observations Survey (ViNOS; J. Licandro et al. 2023) or the European Near Earth Asteroids Research (EURONEAR; O. Vaduvescu et al. 2008). On the other hand, sparse data refers to a dataset consisting of observations taken on separate nights, with only a few measurements per night, typically distributed over long time spans of months or even years. Examples of sparse data are those obtained by ATLAS and NEOWISE. It is possible to use only dense light curves (J. Torppa et al. 2003; J. Ďurech et al. 2007), only sparse (J. Ďurech et al. 2016, 2019), or a balanced mixture (J. Ďurech et al. 2009). In the last case, dense light

---

[9] https://cneos.jpl.nasa.gov/about/neo_groups.html
[10] https://cneos.jpl.nasa.gov/stats/totals.html
[11] See the following webpage for a list of discoveries per survey: https://www.minorplanetcenter.net/iau/lists/MPDiscsNum.html.

**Table 1**
Summary of Prior Published Physical Information for the Four NEAs Analyzed in This Work

| Object | $H$ | Albedo | Diameter (km) | Rot. Period (hr) | LC Amplitude (mag) | Taxonomy | PHA? |
|---|---|---|---|---|---|---|---|
| (21088) Chelyabinsk | 14.33 | 0.179 | 2.79 | 22.48 | 0.13 | S/Sq/Q | N |
| (66146) 1998 TU₃ | 14.41 | 0.224–0.329 | 3.31 | 2.378 | 0.12–0.15 | S/Q | N |
| (297418) 2000 SP₄₃ | 18.52 | 0.38–0.39 | 0.44 | 6.314 | 0.98–1.13 | S/V | Y |
| (4660) Nereus | 18.30 | 0.39 | 0.51/0.33/0.24 | 15.16 | 0.66–0.75 | C/E/X/Xe | Y |

**Note.** Absolute magnitudes $H$, geometric albedo, diameters, rotation periods, light-curve amplitudes, and taxonomic classifications are taken from the Near-Earth Objects Coordination Centre (https://neo.ssa.esa.int/search-for-asteroids) physical properties database.

curves always have a higher weight in the code than sparse data.

In this work only dense light curves are used, since the new observations we obtained in the framework of ViNOS and the published data obtained from the Asteroid Lightcurve Data Exchange Format (ALCDEF;[12] R. D. Stephens et al. 2010; B. D. Warner et al. 2011; R. Stephens & B. D. Warner 2018) public database belong to this type. The main goal of the ViNOS program is to characterize NEAs by using spectroscopic, spectrophotometric, and light-curve observations using telescopes in the Canary Islands observatories (El Roque de los Muchachos Observatory and the Teide Observatory in the La Palma and Tenerife islands, respectively). The program focuses on recently discovered NEAs, those having the smallest sizes, those classified as potentially hazardous (PHAs), possible targets for space missions, and those observed with other techniques (like radar), to provide complementary data. In particular, we have obtained dense light curves of more than 200 NEAs since 2018, with the aim of studying their rotational properties and shapes.

It is important to highlight two effects that can influence NEAs and play a crucial role: one is the Yarkovsky effect (I. Yarkovsky 1901; W. F. J. Bottke et al. 2006; D. Vokrouhlický et al. 2015), and the other is the Yarkovsky–O'Keefe–Radzievskii–Paddack (YORP; I. Yarkovsky 1901; V. V. Radzievskii 1952; S. J. Paddack 1969; J. O'Keefe 1976; W. F. J. Bottke et al. 2006; D. Vokrouhlický et al. 2015) effect. The Yarkovsky effect is caused by the thermal reemission of solar radiation absorbed by an asteroid's surface, which leads to a gradual change in its orbital semimajor axis over time—an increase for prograde rotators and a decrease for retrograde ones. On the other hand, the YORP effect may influence the spin properties, changing the rotation period of the asteroid over time owing to the anisotropy of these reemissions of the thermal radiation. In some cases, if the YORP effect is strong enough, there is evidence that even the spin poles can change over a sufficient amount of time (J. Ďurech et al. 2008a, 2024). Both effects are sensitive to thermal inertia, surface roughness, and albedo. Measuring them helps constrain how the surface stores and reemits heat, offering clues about composition and regolith characteristics.

In this work, we present 61 new light curves, corresponding to ~280 hr of telescope time, for four NEAs: 38 for (21088) Chelyabinsk, 2 for (66146) 1998 TU3, 13 for (4660) Nereus, and 8 for (297418) 2000 SP43, the last two being PHAs. In addition to the new observations, we also used archival data from online databases: 7 for (21088) Chelyabinsk, 36 for (66146) 1998 TU3, 8 for (4660) Nereus, and 5 for (297418) 2000 SP43. Prior published physical information for the four

NEAs is shown in Table 1. Each target of the sample analyzed in this work was selected based on its specific scientific interest and its accessibility within the ViNOS observing schedule. Two of them, (4660) Nereus and (297418) 2000 SP43, are PHAs and thus directly relevant for planetary defense. (66146) 1998 TU3 is a possible YORP candidate, a phenomenon rarely confirmed observationally, while (21088) Chelyabinsk presents conflicting period determinations in the literature that our new data help to resolve. Moreover, this study represents part of a long-term ViNOS effort to expand the number of NEAs with well-constrained spin and shape models, contributing to the statistical characterization of their rotational properties. Notice that to obtain good shapes and poles from light-curve inversion techniques a large amount of good signal-to-noise ratio (S/N) light curves, obtained at different epochs, are needed. Only a small fraction of the NEAs observed in ViNOS have enough data (in particular, previous available good S/N data) to do it. In any case, an ongoing program like ViNOS will continue to be a significant source of NEA light curves for future modeling efforts.

This work is structured as follows: in Section 2, the observational circumstances, instruments used, and reduction methods applied are described. Section 3 describes the methodology used to obtain the shape models and the spin parameters. In Section 4, we present the workflow and the results for each asteroid. In Section 5, we summarize the results of the work.

## 2. Observations

Images of asteroids (21088) Chelyabinsk, (66146) 1998 TU3, (4660) Nereus, and (297418) 2000 SP43 were obtained using four different telescopes, the 0.8 m TTT1, the 0.8 m IAC80 telescope, and the 0.46 m TAR2, all at Teide Observatory (Tenerife, Spain), and the 0.35 m telescope (Isaac Aznar Observatory, Spain). The observing circumstances of each observing run are presented in Table 2. The integration time for each image was selected each night taking into account the asteroid's apparent velocity and visual magnitude, with the aim of obtaining the best possible S/N. The asteroid proper motion, geometric distance, and phase angle were obtained from the Minor Planet Center, while latitude and longitude phase-angle bisectors (PABs) were extracted from the JPL HORIZON web interface. The PAB is the direction that points midway between the asteroid–Earth and asteroid–Sun directions, as it would be seen by an observer located on the asteroid.

The Two-meter Twin Telescope facility (TTT) is located at the Teide Observatory (latitude: 28°18′ 01″.8 N; longitude: +16° 30′ 39″.2 W; altitude: 2386.75 m), on the island of Tenerife (Canary Islands, Spain). Currently, it includes two

**Table 2**
Observational Circumstances of the New Light Curves Acquired by ViNOS

| Asteroid | Telescope | Filter | Exp. Time | Date | UT (Start) | UT (End) | $\alpha$ (deg) | $r$ (au) | $\Delta$ (au) | PABLon (deg) | PABLat (deg) |
|---|---|---|---|---|---|---|---|---|---|---|---|
| 4660 Nereus (1982 DB) | TAR2 | V | 180 | 2021 Oct 3 | 01:13:16.637 | 6:05:26.333 | 26.63 | 1.3222 | 0.3755 | 32.56 | 3.01 |
| 4660 Nereus (1982 DB) | TAR2 | V | 180 | 2021 Oct 4 | 22:38:12.221 | 6:10:01.430 | 25.89 | 1.3119 | 0.3603 | 33.40 | 3.12 |
| 4660 Nereus (1982 DB) | TAR2 | V | 180 | 2021 Oct 05 | 22:18:12.125 | 6:11:07.354 | 25.50 | 1.3066 | 0.3526 | 33.84 | 3.18 |
| 4660 Nereus (1982 DB) | TAR2 | V | 120 | 2021 Oct 14 | 22:04:25.363 | 6:20:25.584 | 21.53 | 1.2577 | 0.2859 | 37.86 | 3.76 |
| 4660 Nereus (1982 DB) | TAR2 | V | 120 | 2021 Oct 15 | 22:04:28.819 | 6:15:57.917 | 21.06 | 1.2523 | 0.2789 | 38.31 | 3.84 |
| 4660 Nereus (1982 DB) | TAR2 | V | 180 | 2021 Oct 28 | 20:23:08.419 | 4:52:11.424 | 14.79 | 1.1832 | 0.1975 | 44.34 | 5.02 |
| 4660 Nereus (1982 DB) | TAR3 | V | 70 | 2021 Oct 28 | 23:16:43.075 | 0:39:04.550 | 14.74 | 1.1825 | 0.1968 | 44.39 | 5.03 |
| 4660 Nereus (1982 DB) | TAR3 | V | 140 | 2021 Oct 30 | 20:43:55.690 | 1:10:32.131 | 13.92 | 1.1726 | 0.1862 | 45.32 | 5.25 |
| 4660 Nereus (1982 DB) | TAR2 | V | 30 | 2021 Nov 30 | 21:36:03.629 | 3:37:02.784 | 27.53 | 1.0292 | 0.0487 | 68.09 | 15.06 |
| 4660 Nereus (1982 DB) | TAR2 | V | 30 | 2021 Nov 30 | 23:50:32.784 | 2:55:17.040 | 27.70 | 1.0288 | 0.0484 | 68.22 | 15.15 |
| 4660 Nereus (1982 DB) | TAR3 | V | 13 | 2021 Dec 02 | 21:24:55.670 | 0:00:31.622 | 31.50 | 1.0218 | 0.0424 | 71.48 | 17.01 |
| 4660 Nereus (1982 DB) | TAR3 | V | 10 | 2021 Dec 08 | 01:23:00.960 | 6:03:46.282 | 49.52 | 1.0040 | 0.0295 | 85.85 | 24.19 |
| 4660 Nereus (1982 DB) | TAR2 | V | 10 | 2021 Dec 18 | 04:14:42.950 | 6:56:21.782 | 101.67 | 0.9761 | 0.0361 | 134.98 | 23.58 |
| 21088 Chelyabinsk (1992 BL2) | TAR2 | V | 90 | 2022 Nov 23 | 00:03:06.278 | 6:29:54.182 | 30.15 | 1.7432 | 1.0502 | 92.84 | 36.57 |
| 21088 Chelyabinsk (1992 BL2) | TAR2 | V | 120 | 2022 Nov 29 | 00:12:34.704 | 6:28:37.459 | 28.14 | 1.7614 | 1.0210 | 93.07 | 35.47 |
| 21088 Chelyabinsk (1992 BL2) | TAR2 | V | 120 | 2022 Dec 1 | 00:12:38.938 | 6:28:41.606 | 27.43 | 1.7674 | 1.0122 | 93.08 | 35.07 |
| 21088 Chelyabinsk (1992 BL2) | T35/Isaac Aznar | Clear | 60 | 2024 Jul 10 | 23:28:42.528 | 2:23:17.664 | 30.85 | 1.3230 | 0.3785 | 315.05 | R 3.53 |
| 21088 Chelyabinsk (1992 BL2) | T35/Isaac Aznar | Clear | 60 | 2024 Jul 11 | 23:56:30.912 | 2:10:03.648 | 30.30 | 1.3216 | 0.3736 | 315.25 | R 4.63 |
| 21088 Chelyabinsk (1992 BL2) | T35/Isaac Aznar | Clear | 60 | 2024 Jul 15 | 23:32:29.760 | 2:11:11.040 | 28.55 | 1.3167 | 0.3583 | 315.89 | R 9.06 |
| 21088 Chelyabinsk (1992 BL2) | T35/Isaac Aznar | Clear | 30 | 2024 Jul 23 | 21:34:55.200 | 3:40:09.408 | 28.42 | 1.3087 | 0.3484 | 316.53 | R 18.14 |
| 21088 Chelyabinsk (1992 BL2) | T35/Isaac Aznar | Clear | 30 | 2024 Jul 23 | 23:48:36.576 | 3:00:34.560 | 28.45 | 1.3086 | 0.3485 | 316.53 | R 18.24 |
| 21088 Chelyabinsk (1992 BL2) | T35/Isaac Aznar | Clear | 30 | 2024 Jul 30 | 20:26:47.616 | 3:39:17.568 | 32.10 | 1.3037 | 0.3623 | 316.67 | R 25.58 |
| 21088 Chelyabinsk (1992 BL2) | T35/Isaac Aznar | Clear | 30 | 2024 Jul 30 | 23:56:17.952 | 3:33:38.304 | 32.20 | 1.3036 | 0.3628 | 316.67 | R 25.73 |
| 21088 Chelyabinsk (1992 BL2) | T35/Isaac Aznar | Clear | 30 | 2024 Jul 31 | 21:08:43.584 | 0:44:35.808 | 32.81 | 1.3031 | 0.3660 | 316.69 | R 26.60 |
| 21088 Chelyabinsk (1992 BL2) | T35/Isaac Aznar | Clear | 30 | 2024 Aug 1 | 00:51:08.064 | 4:14:55.392 | 32.91 | 1.3030 | 0.3665 | 316.69 | R 26.75 |
| 21088 Chelyabinsk (1992 BL2) | T35/Isaac Aznar | Clear | 30 | 2024 Aug 2 | 00:57:47.232 | 4:15:30.816 | 33.62 | 1.3025 | 0.3705 | 316.71 | R 27.71 |
| 21088 Chelyabinsk (1992 BL2) | T35/Isaac Aznar | Clear | 30 | 2024 Aug 2 | 22:38:37.536 | 4:09:34.934 | 34.26 | 1.3021 | 0.3744 | 316.73 | R 28.55 |
| 21088 Chelyabinsk (1992 BL2) | TTT1 | L | 50 | 2024 Aug 15 | 20:38:28.406 | 0:48:17.510 | 42.36 | 1.2997 | 0.4520 | 317.80 | 38.37 |
| 21088 Chelyabinsk (1992 BL2) | TTT1 | L | 50 | 2024 Aug 17 | 20:37:47.021 | 0:33:35.194 | 43.29 | 1.3000 | 0.4665 | 318.14 | 39.56 |
| 21088 Chelyabinsk (1992 BL2) | TTT1 | L | 50 | 2024 Aug 21 | 21:13:09.782 | 0:28:41.606 | 44.86 | 1.3010 | 0.4968 | 318.99 | 41.74 |
| 21088 Chelyabinsk (1992 BL2) | TTT1 | L | 50 | 2024 Aug 22 | 20:30:23.357 | 0:20:18.154 | 45.18 | 1.3014 | 0.5043 | 319.22 | 42.23 |
| 21088 Chelyabinsk (1992 BL2) | TTT1 | L | 60 | 2024 Aug 23 | 21:57:43.949 | 0:18:17.453 | 45.52 | 1.3018 | 0.5125 | 319.49 | 42.75 |
| 21088 Chelyabinsk (1992 BL2) | TTT1 | L | 70 | 2024 Aug 24 | 21:54:15.725 | 0:03:12.067 | 45.81 | 1.3022 | 0.5203 | 319.76 | 43.22 |
| 21088 Chelyabinsk (1992 BL2) | TTT1 | L | 60 | 2024 Aug 25 | 20:44:44.419 | 3:59:48.422 | 46.06 | 1.3027 | 0.5278 | 320.03 | 43.66 |
| 21088 Chelyabinsk (1992 BL2) | TTT1 | L | 60 | 2024 Aug 26 | 20:27:09.821 | 0:00:08.899 | 46.31 | 1.3032 | 0.5356 | 320.31 | 44.10 |
| 21088 Chelyabinsk (1992 BL2) | TTT1 | L | 70 | 2024 Aug 29 | 20:20:32.381 | 3:09:46.109 | 46.96 | 1.3050 | 0.5592 | 321.26 | 45.37 |
| 21088 Chelyabinsk (1992 BL2) | TTT1 | L | 80 | 2024 Aug 30 | 20:21:02.102 | 2:38:57.754 | 47.15 | 1.3057 | 0.5671 | 321.59 | 45.77 |
| 21088 Chelyabinsk (1992 BL2) | TTT1 | L | 80 | 2024 Aug 31 | 20:17:29.386 | 3:44:04.934 | 47.31 | 1.3064 | 0.5750 | 321.94 | 46.16 |
| 21088 Chelyabinsk (1992 BL2) | TTT1 | L | 80 | 2024 Sep 1 | 20:24:37.238 | 3:38:26.246 | 47.47 | 1.3072 | 0.5830 | 322.30 | 46.54 |
| 21088 Chelyabinsk (1992 BL2) | TTT1 | L | 90 | 2024 Sep 5 | 20:13:31.526 | 0:25:57.850 | 47.93 | 1.3106 | 0.6143 | 323.85 | 47.95 |
| 21088 Chelyabinsk (1992 BL2) | TTT1 | L | 100 | 2024 Sep 6 | 20:10:26.198 | 0:48:37.008 | 48.02 | 1.3116 | 0.6221 | 324.26 | 48.28 |
| 21088 Chelyabinsk (1992 BL2) | TTT1 | L | 120 | 2024 Sep 8 | 20:58:44.314 | 1:54:56.851 | 48.16 | 1.3137 | 0.6379 | 325.13 | 48.93 |
| 21088 Chelyabinsk (1992 BL2) | TTT1 | L | 110 | 2024 Sep 9 | 20:22:42.413 | 1:53:21.379 | 48.21 | 1.3147 | 0.6454 | 325.57 | 49.22 |
| 21088 Chelyabinsk (1992 BL2) | TTT1 | L | 210 | 2024 Sep 13 | 00:08:08.246 | 1:06:37.037 | 48.32 | 1.3184 | 0.6694 | 327.02 | 50.13 |
| 21088 Chelyabinsk (1992 BL2) | TTT1 | L | 200 | 2024 Sep 15 | 01:01:25.565 | 1:55:17.011 | 48.34 | 1.3210 | 0.6846 | 328.01 | 50.68 |

**Table 2**
(Continued)

| Asteroid | Telescope | Filter | Exp. Time | Date | UT (Start) | UT (End) | $\alpha$ (deg) | $r$ (au) | $\Delta$ (au) | PABLon (deg) | PABLat (deg) |
|---|---|---|---|---|---|---|---|---|---|---|---|
| 21088 Chelyabinsk (1992 BL2) | TTT1 | $L$ | 170 | 2024 Sep 15 | 23:49:14.938 | 0:34:47.078 | 48.34 | 1.3222 | 0.6917 | 328.49 | 50.93 |
| 21088 Chelyabinsk (1992 BL2) | TTT1 | $L$ | 200 | 2024 Sep 20 | 22:22:30.634 | 1:16:10.387 | 48.24 | 1.3292 | 0.7272 | 331.07 | 52.10 |
| 21088 Chelyabinsk (1992 BL2) | TTT1 | $L$ | 180 | 2024 Sep 21 | 22:21:09.936 | 0:30:30.384 | 48.20 | 1.3307 | 0.7342 | 331.62 | 52.31 |
| 21088 Chelyabinsk (1992 BL2) | TTT1 | $L$ | 200 | 2024 Sep 22 | 22:22:08.947 | 0:28:57.418 | 48.16 | 1.3323 | 0.7411 | 332.17 | 52.52 |
| 21088 Chelyabinsk (1992 BL2) | TTT1 | $L$ | 200 | 2024 Sep 23 | 23:12:34.243 | 0:03:47.578 | 48.11 | 1.3340 | 0.7483 | 332.75 | 52.73 |
| 21088 Chelyabinsk (1992 BL2) | TTT1 | $L$ | 280 | 2024 Sep 25 | 00:25:15.370 | 0:44:35.117 | 48.06 | 1.3357 | 0.7554 | 333.35 | 52.94 |
| 66146 (1998 TU3) | IAC80 | $r$ | 45 | 2022 Sep 6 | 03:42:06.595 | 5:53:04.675 | 61.35 | 1.1213 | 0.7571 | 54.97 | −7.64 |
| 66146 (1998 TU3) | TAR2 | $V$ | 60 | 2022 Sep 8 | 03:14:22.272 | 5:32:18.355 | 61.95 | 1.1148 | 0.7418 | 56.49 | −7.69 |
| 297418 (2000 SP43) | TAR2 | $V$ | 60 | 2022 Oct 18 | 20:02:40.675 | 3:27:31.853 | 34.12 | 1.1861 | 0.2403 | 39.20 | 20.66 |
| 297418 (2000 SP43) | TAR2 | $V$ | 60 | 2022 Oct 19 | 20:31:37.229 | 3:26:33.187 | 33.45 | 1.1870 | 0.2393 | 38.76 | 20.96 |
| 297418 (2000 SP43) | TAR2 | $V$ | 60 | 2022 Oct 20 | 20:00:40.925 | 3:25:42.038 | 32.88 | 1.1877 | 0.2386 | 38.32 | 21.25 |
| 297418 (2000 SP43) | TAR2 | $V$ | 50 | 2022 Nov 1 | 19:06:31.248 | 3:56:22.790 | 33.16 | 1.1893 | 0.2460 | 32.62 | 23.56 |
| 297418 (2000 SP43) | TAR2 | $V$ | 50 | 2022 Nov 2 | 19:06:22.090 | 0:56:17.606 | 33.74 | 1.1888 | 0.2479 | 32.19 | 23.66 |
| 297418 (2000 SP43) | TAR2 | $V$ | 50 | 2022 Nov 3 | 19:06:24.768 | 0:56:25.296 | 34.39 | 1.1882 | 0.2500 | 31.77 | 23.74 |
| 297418 (2000 SP43) | TAR2 | $V$ | 50 | 2022 Nov 4 | 19:06:26.410 | 0:56:27.024 | 35.09 | 1.1875 | 0.2523 | 31.36 | 23.80 |
| 297418 (2000 SP43) | TAR2 | $V$ | 50 | 2022 Nov 16 | 19:35:16.397 | 3:57:04.262 | 45.28 | 1.1715 | 0.2903 | 28.10 | 23.80 |

**Note.** The table includes the asteroid IDs, telescopes used, filters used ($r$ SLOAN, $V$, Clear, and Luminance), the date, the starting and end times (UT) of the observations, the phase angle ($\alpha$), the heliocentric ($r$) and geocentric ($\Delta$) distances, and the phase-angle bisector longitude (PABLon) and latitude (PABLat) of the asteroid at the time of observation.

0.8 m telescopes (TTT1 and TTT2) and a 2.0 m telescope (TTT3) on altazimuth mounts. We used the TTT1 telescope, which has two Nasmyth ports with focal ratios of $f/D =$ 6.8 and $f/D = 4.4$ equipped with QHY411M[13] CMOS cameras (M. R. Alarcon et al. 2023). The QHY411M has scientific Complementary Metal–Oxide–Semiconductor (sCMOS) image sensors with 14 K $\times$ 10 K 3.76 $\mu$m pixel$^{-1}$ pixels. This setup, in the $f/D = 4.4$ focus, provides an effective field of view of $51.\!'4 \times 38.\!'3$ (with an angular resolution of $0.\!''22$ pixel$^{-1}$). Images were taken using the *Luminance* filter, which covers the 0.4–0.72 $\mu$m wavelength range, and the exposure time was dynamically set between to ensure an S/N higher than 50. Images were bias and flat-field corrected in the standard way.

The IAC80 is a 82 cm telescope with an $f/D = 11.3$ in the Cassegrain focus. It is equipped with the CAMELOT-2 camera, a back-illuminated e2v 4 K $\times$ 4 K pixel CCD of 15 $\mu$m pixels. This setup provide a plate scale of $0.\!''32$ pixel$^{-1}$ and a field of view of $21.\!'98 \times 22.\!'06$. Images were obtained using the Sloan *r* filter with the telescope in sidereal tracking, so the individual exposure time of the images was selected such that the asteroid trail was smaller than the typical FWHM of the IAC80 images ($\sim$1.''0). The images were bias and flat-field corrected in the standard way.

TAR2 is a 46 cm robotic telescope with $f/D = 2.8$ in the prime focus (actually decommissioned). TAR2 was equipped with a QHY600M-PRO camera, a back-illuminated 9K $\times$ 6K sCMOS of the same kind as that of the QHY411M used by the TTT1 telescope, with a pixel size of 3.76 $\mu$m, binned by a 2 $\times$ 2 factor that provides a $1.\!''22$ pixel$^{-1}$ scale. The images were bias and flat-field corrected in the standard way.

The Isaac Aznar/T35 telescope is a Schmidt–Cassegrain 14" telescope installed in a private observatory in Valencia (Spain) equipped with an STL-1001E 1K $\times$ 1K CCD camera that has a pixel scale of $1.\!''44$ pixel$^{-1}$. The images were bias, dark, and flat-field corrected in the standard way.

In the case of TTT1, IAC80, and TAR2 images, to obtain the light curves, we performed aperture photometry of the final images using the Photometry Pipeline[14] (PP) (M. Mommert 2017), as we did in J. Licandro et al. (2023). The images obtained with the *L* filter were calibrated to the *r* SLOAN band using the Pan-STARRS catalog, while the other images were calibrated to the corresponding bands for the filters used. Data obtained with the Isaac Aznar/T35 were reduced using MPO Canopus V10, and differential aperture photometry was obtained and calibrated using field stars with magnitudes obtained from the photometric reference catalog MPOSC3 as in A. Aznar et al. (2019). Although all observations were carried out using a clear filter in order to maximize the asteroid light flux signal, *V* magnitudes of comparison stars were selected to obtain the calibrated *V* magnitude of the asteroid.

## 3. Light-curve Inversion Method

In this work, we follow same light-curve inversion methodology as described in J. Rodríguez Rodríguez et al. (2024a, 2024b), using two different inversion codes. The first, referred to here as the no-YORP code, is publicly available through the Database of Asteroid Models from Inversion Techniques (DAMIT;[15] J. Ďurech et al. 2010) and assumes a constant rotation period, allowing for the derivation of shape models and spin axis orientations. The second, referred to as the YORP code, extends this approach by allowing the rotation period to vary linearly with time, thereby accounting for the YORP effect. This code is not publicly available and was kindly provided by Josef Ďurech via personal communication. Only models with spin axis aligned with the short principal axis are considered.

Some key parameters are needed to fully characterize an asteroid and create a shape model with both codes. The first parameter is the sidereal rotation period ($P$), which is the time the asteroid takes to complete a single rotation over its spin axis with respect to the background stars and is needed as the initial parameter for both codes. To obtain this value, we first checked for previous publications in ALCDEF (see footnote 12) to obtain an initial range to run the period scan tool provided with the no-YORP code. For each asteroid we performed a search in a wide margin that covers the values in ALCDEF, and then we proceeded to do a fine search around the best fit to the data in this work. This tool uses a $\chi^2$ cost function to obtain the period that best fits the light curves provided. The inputs for this code are the interval of periods in which the search will be carried out, the period step coefficient ($p$), and the convexity regularization weight ($d$), which helps to maintain the dark facet area below the threshold of 1%. Our first search takes into account all the $P$ published values for a given asteroid, finding with it a global minimum; subsequent searches are made tightening the interval to refine the period until no improvement in the fit is made.

Once an initial value for $P$ is obtained, we proceed to run the no-YORP code, which has as initial parameters $P$, $\lambda$, and $\beta$, with $\lambda$ and $\beta$ being the ecliptic coordinates of the spin axis (ecliptic longitude and latitude, respectively); their values are in the ranges $0° \leqslant \lambda \leqslant 360°$ and $-90° \leqslant \beta \leqslant 90°$. This code has additional parameters, such as the convexity regularization weight ($d$), which would be modified as needed after the first search if the value for $d$ for the best medium solution is >1%; the light scattering parameters and Lambert coefficient, which will be left as default, amplitude $a = 0.5$, width $d = 0.1$, slope $k = -0.5$, and $c = 0.1$, respectively; while the number of iterations is set at 50 for each pair of $\lambda$ and $\beta$. The base code works by finding a $\chi^2$ for a given set of the previously mentioned values, so we upgraded it by iterating the values of $\lambda$ and $\beta$ in steps of 5° for the entire sphere ($\sim$ 2700 poles), leaving $P$ as a free value, that is, $P$ can be adjusted by the code, while $\lambda$ and $\beta$ are fixed to their initial values.

Once the medium search is computed, we proceed to reduce the $\chi^2$ to the number of observations following Equation (1), thus obtaining $\chi^2_{red}$:

$$\chi^2_{red} = \frac{\chi^2}{\nu}. \tag{1}$$

In Equation (1) $\nu$ is the number of degrees of freedom, which is the number of observations for the asteroids (number of measures per light curve) minus the number of parameters ($\sim$100; D. Vokrouhlický et al. 2011). The advantage of a $\chi^2_{red}$ is that a value near 1 means that the model perfectly fits the data, which is usually impossible owing to the data and their uncertainties. The lowest $\chi^2_{red}$ solution would be the starting

---

[13] https://www.qhyccd.com/

[14] https://photometrypipeline.readthedocs.io/en/latest/

[15] https://astro.troja.mff.cuni.cz/projects/damit/

**Table 3**
The Obtained Rotation Period ($P$), Geocentric Ecliptic Coordinates of the Spin Pole ($\lambda$, $\beta$), Obliquity ($\epsilon$), a/b and b/c Aspect Ratios, and YORP Acceleration ($v$, If Determined)

| Asteroid | Model | Period (hr) | $\lambda$ (deg) | $\beta$ (deg) | $\epsilon$ (deg) | a/b | b/c | $v$ (rad day$^{-2}$) |
|---|---|---|---|---|---|---|---|---|
| (4660) Nereus | C | 15.159442 ± 0.000738 | 321 ± 19 | 78 ± 12 | 12 ± 11 | 1.938 | 2.123 | ⋯ |
| (21088) Chelyabinsk | C | 11.227651 ± 0.000025 | 232 ± 2 | −55 ± 1 | 108 ± 1 | 1.197 | 1.668 | ⋯ |
| (66146) 1998 TU3 | C | 2.377471 ± 0.000001 | 64 ± 6 | −70 ± 1 | 157 ± 1 | 1.099 | 1.532 | ⋯ |
| (66146) 1998 TU3 | L | 2.377473 ± 0.000002 | 64 ± 5 | −71 ± 2 | 157 ± 1 | 1.128 | 1.366 | (2.37 ± 2.09) ×10$^{-8}$ |
| (297418) 2000 SP3 | C | 6.312233 ± 0.000022 | 265 ± 5 | 54 ± 4 | 26 ± 4 | 1.778 | 6.853 | ⋯ |

**Note.** Depending on the dataset, the linearly increasing rotation period (L) or the constant rotation period model (C) is applied. Notice that Nereus and 2000 SP43 Yarkovsky accelerations in the Small-Body Database match our determined spin pole directions.

point for both the fine solution and the subsequent uncertainties, which will be discussed later. In addition, this search is the one plotted in Appendix B for each model. For this pole plot, a threshold of 10% of the lowest $\chi^2_{red}$ is used as the solutions that have the best fit with the data.

For the fine search we proceed by reducing the search to a $30° \times 30°$ square around the lowest $\chi^2_{red}$ medium solution in 2° steps, taking this time the $P$ of this medium solution, leaving $\lambda$ and $\beta$ also free (the code will adjust them to their best-fitting values), and obtaining this time ~250 poles. Carrying the computation in this particular way, almost all the fine solutions should converge if the data cover enough time and different solar phase angles $\alpha$ and aspect angles $\theta$ of the object. As for the medium search, we also proceed to obtain the $\chi^2_{red}$, for the fine solutions, adopting the lowest one to create the shape model with the model creation tool to create the shape model that is provided with each asteroid. The models created with the model creation tool are used to compute synthetic light curves, which are those plotted in Sections 4.2, 4.3, and 4.4 and Appendix C. To fully characterize the spin axis for a given fine solution, the obliquity ($\epsilon$) needs to be calculated. For this, we obtained the values of the asteroid's inclination (i), longitude of ascending node ($\Omega$), and argument of pericenter ($\omega$) from the Horizons System (J. D. Giorgini et al. 1996, 2001), which, along with the asteroid fine solution, allows us to calculate $\epsilon$. This value helps us to know the spin pointing direction of the asteroid, given that if $0° \leqslant \epsilon \leqslant 90°$ the asteroid would be a prograde rotator, while if $90° < \epsilon \leqslant 180°$ it would be retrograde.

The last step is the computation of the uncertainties, which we decided to approach as follows: First, we created 100 sets of light curves from the initial one, but randomly removing 25% of the data for each light curve. After that, for each set, we computed a fine search, thus obtaining 100 different fine solutions. With these 100 values, we computed the mean (which is almost identical to the fine solution with the initial data) and the standard deviation (3$\sigma$), which will serve as our solution uncertainty.

This process is exactly the same for the "YORP code," but in this case another parameter is present, $v$, which is the value of the YORP acceleration. This value is set, in both the medium and fine searches, at $v = 1 \times 10^{-8}$ rad day$^{-2}$ and set free, allowing it to be adjusted to its best-fit value. If a search is carried out with this code, as a way to validate our conclusion, another method is used, the one proposed in D. Vokrouhlický et al. (2017), which yields a value for $v$ and its uncertainties at the 3$\sigma$ level. This validation consists in performing a search fixing all the values at the fine solution except $v$, which will be iterated in a range according to the fine solution. It is worth mentioning that this validation usually verifies our previous solution.

## 4. Results

The results obtained using the methodology described in Section 3 are presented in Table 3. In Sections 4.1, 4.2, 4.3, and 4.4, we detail the analysis process for each asteroid and discuss the corresponding results.

### 4.1. (4660) Nereus

This is an object belonging to the Apollo group, which is characterized by a semimajor axis ($a$) >1.0 au and a perihelion distance ($q$) <1.017 au, as per CNEOS (see footnote 9). It is also tagged as PHA owing to its Earth MOID of 0.00367 au. There are two published diameters, one from thermal infrared data (0.33 km; M. Delbó et al. 2003) and another from radar data (0.33 ± 0.03 km; M. Brozovic et al. 2009).

We digitalized the two light curves from Y. Ishibashi et al. (2000), which cover the second and third nights of 1997 August; from the ALCDEF database we obtained six light curves from 2021 October 26 to 2021 November 3; and from the ViNOS project another 13 light curves from 2021 October 3 to 2021 December 18 were added, for a total of 21 light curves and a temporal span of ~24 yr. For this asteroid, there is a published model from radar data taken in 2002 (M. Brozovic et al. 2009), with a spin axis of $\lambda = 25° \pm 10°$ and $\beta = 80° \pm 10°$; it is also suggested that the asteroid may be affected by YORP (M. Brozovic et al. 2009).

There are several published periods for this object: $P = 15.1$ hr (Y. Ishibashi et al. 1998, 2000), $P = 15.16 \pm 0.04$ hr (M. Brozovic et al. 2009), $P = 15.172 \pm 0.002$ hr (B. D. Warner & R. D. Stephens 2022), and $P = 15.26 \pm 0.01$ hr (L. Franco et al. 2022). We took all those values into account for the period tool, obtaining as the best-fitting value $P = 15.156877$ hr, as shown in Figure 1.

We then proceed to compute the medium search with that period as a starting point, obtaining $P = 15.158726$ hr, $\lambda = 320°$, $\beta = 80°$, $\chi^2_{red} = 1.48$ (in Appendix B.1 the graphical representation of the distribution of the solutions is shown). With this medium solution a fine solution is computed, with the following results: $P = 15.159839$ hr, $\lambda = 323°$, $\beta = 81°$, $\chi^2_{red} = 1.47$, and $\epsilon \simeq 9°$, which makes Nereus a prograde rotator. In Figure 2 the best-fitting model shape is shown, while in Figure 3 and Appendix C.1 the fit between data and model is plotted.

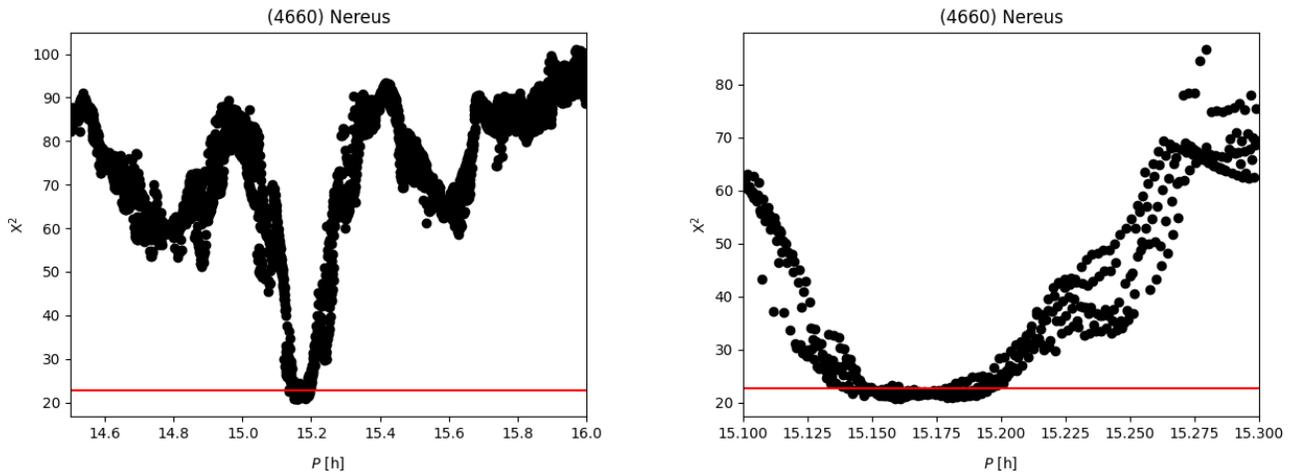

**Figure 1.** Output from the period search tool for (4660) Nereus. Each obtained period is plotted as a black circle, while a solid red line represents a 10% threshold from the lowest $\chi^2$ obtained in the search. Left: a wider interval from 14.5 to 16 hr is used to ensure the minimum of the data presented in this work. Right: the interval used is from 15 to 15.3 hr, with a coefficient $p$ of 0.2. There are several periods under the threshold, which implies that more data are needed to refine the value, while a good amount of them are in the vicinity of 15.15 hr.

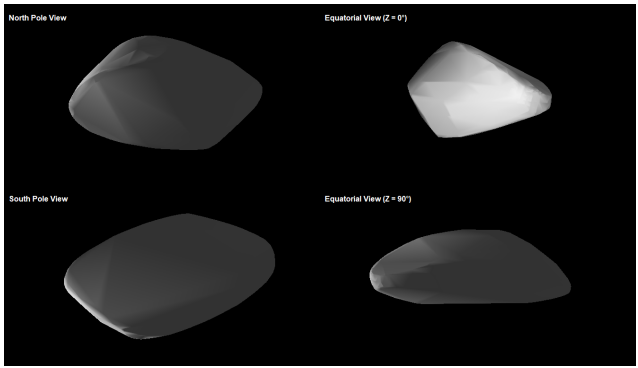

**Figure 2.** Obtained shape model of (4660) Nereus with a constant rotation period. Top left: north pole view ($Y$-axis = 0°). Bottom left: south pole view ($Y$-axis = 180°). Top right: equatorial view with $Z$-axis rotated 0°. Bottom right: equatorial view with $Z$-axis rotated 90°.

Lastly, we compute the uncertainties by creating 100 random subsets, each with 25% of the data removed since it is a relatively large dataset (~5600 points), creating one model for each of them, which is 100 models. Those models are computed taking as a starting point the best medium solution and computing a fine one. We obtain the following: $P = 15.159442 \pm 0.000738$ hr, $\lambda = 321° \pm 19°$, $\beta = 78° \pm 12°$, and $\epsilon = 12° \pm 11°$.

The spin axis presented in this work differs from the one presented in M. Brozovic et al. (2009) in $\lambda$ by a margin ($\lambda = 25°$ from the radar model vs. $\lambda = 323°$ from our model), while $\beta$ is relatively close ($\beta = 80°$ vs. $\beta = 78°$), but that solution is also possible with the data presented in this work, as shown in Appendix B.1. Notice that, due to the high ecliptic latitudes of the two solutions, the angular separation between the corresponding spin vectors is only ~11°.

However, our shape model has a close visual match with both presented in M. Brozovic et al. (2009), as shown in Figure 4, where the same elongated shape is present, on the $X$-axis, and from the top and bottom views ($Z$-axis) the same features can be noticed. In M. Brozovic et al. (2009) it is suggested that (4660) Nereus may be affected by YORP, and while the extreme value of $\epsilon$ of ~12° may be an indicator (J. Hanuš et al. 2013), we were unable to find a model with a better fit with the current data or a value consistent enough to be taken into account.

Following previous studies such as B. D. Warner & R. D. Stephens (2022), where it is claimed that this object is a tumbler, and the discrepancies between the best model and the data (light curves from 2021 October 3 and 2021 October 14), we performed a study of the residuals of the best obtained model and the data. From previous works the secondary period was found to be 12.457 ± 0.02 hr (Pravec 2021web[16]). As shown in Figure 5, the secondary period obtained taking into account all the data is 12.4032 ± 0.0011 hr, while the secondary period taking into account only the data with the lowest S/N (2021 October 26, 27, 28, 29, and 30; 2021 November 3 and 30 and 2021 December 2) is 12.4039 ± 0.0930 hr. These results are in agreement with the previously reported values and could explain the discrepancies between data and the best-fitting model presented here. In any case, it is important to clarify that all shape models, including the Nereus one, assume principal-axis rotation. A complex rotation (non-principal-axis) modeling light-curve inversion model is beyond the scope of this paper.

### 4.2. (21088) Chelyabinsk

This is an object belonging to the Amor group, which is characterized by a semimajor axis ($a$) > 1.0 au and a perihelion distance 1.017 au < ($q$) < 1.3 au, as per CNEOS (see footnote 9). There are four published diameter determinations: $D = 3.46 \pm 0.25$ km (F. Usui et al. 2011), $D = 4.231 \pm 0.113$ km (A. Mainzer et al. 2011b; A. K. Mainzer et al. 2019), $D = 4.232$ km (P. Pravec et al. 2012), and $D = 2.79 \pm 0.10$ km (C. R. Nugent et al. 2016).

A total of 45 light curves, with a temporal span of ~4 yr, are used for the light-curve inversion analysis. In this manuscript we present 38 light curves of (21088) Chelyabinsk, obtained from 2022 November 23 to 2024 September 25 as part of the ViNOS project. Additionally, we used 7 published light curves from the ALCDEF database obtained from 2020 December 15 to 2024 August 3. Since the time span is not large enough, we ran the model assuming a constant rotation period. We also tried to find any signs of YORP acceleration, but there are no clear signs of rotation period variations in our data.

---

[16] https://space.as.cas.cz/~asteroid/04660.png

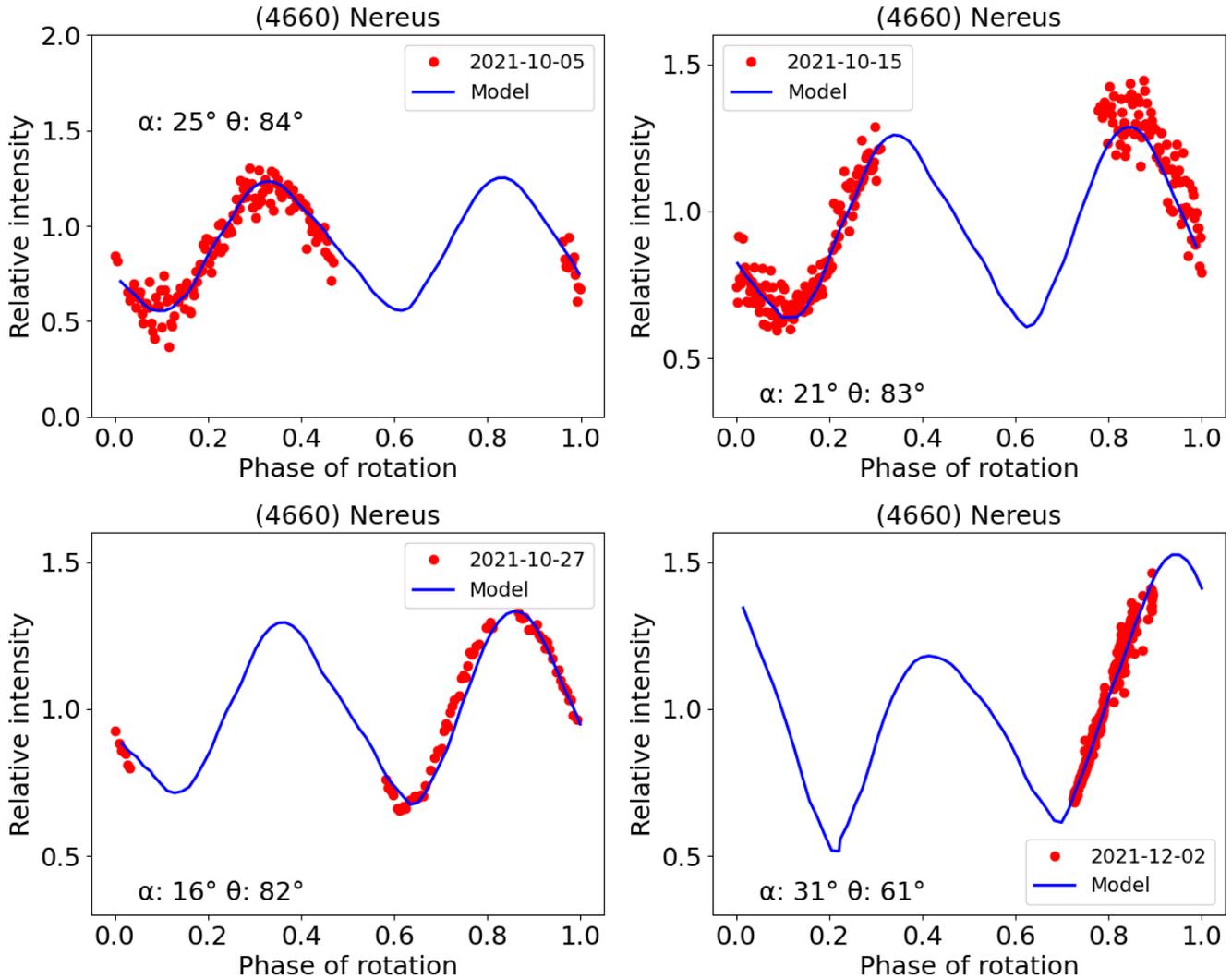

**Figure 3.** Graphical representation of the fit between four of the light curves used in this work (2021 November 3 from ALCDEF and 2021 October 4 and 15 and November 30 from the newly obtained light curves) and the best-fitting constant-period model (C Model) for (4660) Nereus. The data are plotted as red circles for each observation, while the model is plotted as a solid blue line. The geometry is described by its solar phase angle $\alpha$ and its aspect angle $\theta$.

There are several published periods for this object: $P = 22.431 \pm 0.001$ hr and $P = 22.49$ hr (Pravec 2004web[17]), $P = 11.23 \pm 0.01$ hr (B. D. Warner & R. D. Stephens 2021), $P = 11.234 \pm 0.003$ hr (L. Franco et al. 2025), and $P = 11.272 \pm 0.017$ hr (R. G. Farfán et al. 2025).

We took all those values into account and obtained the best-fitting value $P = 11.227649$ hr, as shown in Figure 6. We also tried to detect a period in the neighborhood of 22.43 hr as per Pravec 2004web, but the fitting with the data was always poorer than our solution, as can be seen in Figure 7.

We then proceed to compute the medium search with that period as a starting point, obtaining $P = 11.227645$ hr, $\lambda = 230°$, $\beta = -55°$, and $\chi^2_{\text{red}} = 1.14$ (in Appendix B.2 the graphical representation of the distribution of the solutions is shown). With this medium solution a fine solution is computed, with the following results: $P = 11.227656$ hr, $\lambda = 232°$, $\beta = -55°$, $\chi^2_{\text{red}} = 1.12$, and $\epsilon \simeq 108°$, which makes

Chelyabinsk a retrograde rotator. In Figure 8 the best-fitting model shape is shown, while in Figure 9 and Appendix C.2 the fit between data and model is plotted.

Lastly, we compute the uncertainties by creating 100 random subsets, each with 25% of the data removed since it is a relatively large dataset (~6140 points), creating one model for each of them, which is 100 models. Those models are computed taking as a starting point the best medium solution and computing a fine one. We obtain the following: $P = 11.227651 \pm 0.000025$ hr, $\lambda = 232° \pm 2°$, $\beta = -55° \pm 1°$, and $\epsilon = 108° \pm 1°$.

The shape model of (21088) Chelyabinsk appears highly angular, with large planar facets and sharp edges, giving it a blocky or "brick-like" aspect. This morphology is a well-known and expected outcome of convex light-curve inversion techniques, which favor the simplest convex surface compatible with the photometric data. As a result, models are derived from sparse or moderate aspect angle coverage. A similar, although less pronounced, angular appearance is also observed in the model of (4660) Nereus presented in this work. Similar angular morphologies are observed in other well-established

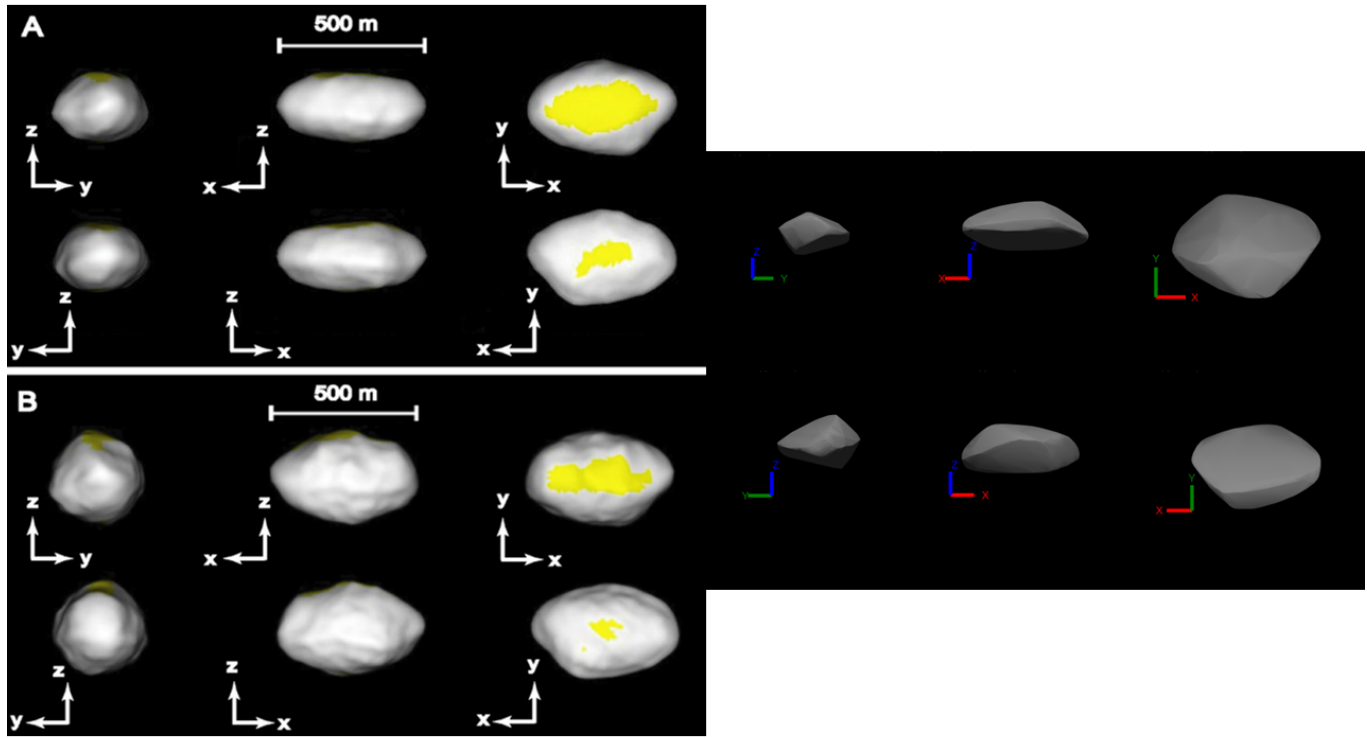

**Figure 4.** Comparison between the radar models proposed in M. Brozovic et al. (2009; left) and our best-fit model (right). The model presented in this work presents the same elongated shape in the *X*-axis and almost the same features in the top and bottom views (seen from the *Z*-axis).

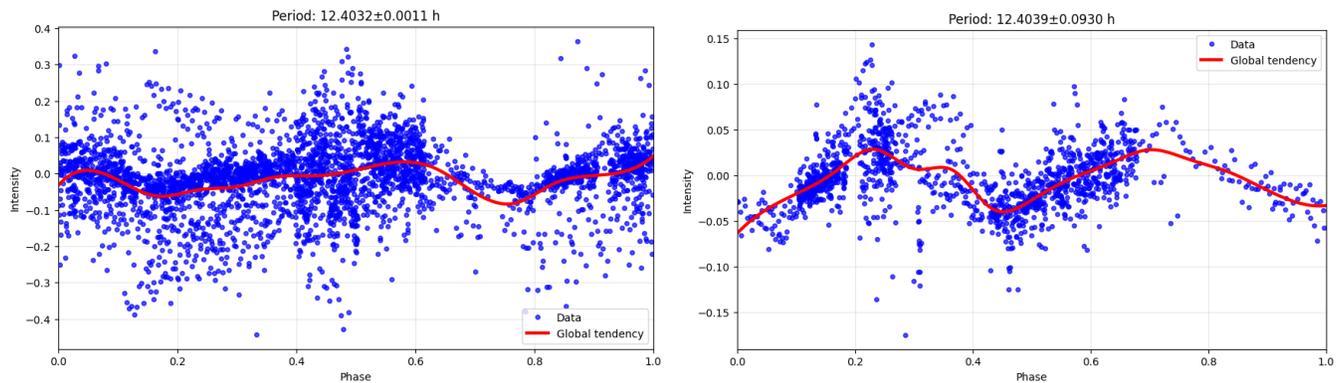

**Figure 5.** Secondary periods obtained from the residuals of the best-fitting model. Left: secondary period taking into account all the light curves presented in this work. Right: secondary period taking into account the light curves with the lowest S/N (see text).

NEA shape models, including the photometric model of (1036) Ganymed[18] (J. Hanuš et al. 2016) and (8567) 1996 HW1,[19] the last one being a contact binary (C. Magri et al. 2011). Therefore, the apparent angularity of the Chelyabinsk model should not be interpreted as an artifact of the inversion process, but rather as a realistic representation of its large-scale shape at the resolution imposed by the available data and the limit of the use of convex-only shapes.

### 4.3. (66146) 1998 TU3

This asteroid belongs to the Atens group, which is characterized by a semimajor axis ($a$) < 1.0 au and an aphelion distance ($Q$) > 0.983 au, as per CNEOS (see footnote 9). The following diameters have been published using WISE data: 2.864 ± 1.165 km (A. Mainzer et al. 2012; A. K. Mainzer et al. 2019), 2.68 ± 0.89 km (J. R. Masiero et al. 2020), and 3.31 ± 1.18 km (J. R. Masiero et al. 2021).

For the light-curve inversion analysis we used a total number of 38 light curves with a temporal span of the data of almost 14 yr. In addition to the 2 light curves presented in this paper, obtained from 2022 September 6 and 8, we used the publicly available 36 light curves obtained from the ALCDEF database that have a temporal span from 2008 October 22 to 2019 October 31, roughly 11 yr.

First, we used the period tool to find the best rotation period, taking as a starting point the published periods: $P = 2.3767 ± 0.0009$ hr and $P = 2.37741 ± 0.00004$ hr (A. Galád et al. 2005), $P = 2.3779 ± 0.0004$ hr (T. Richards et al. 2007), $P = 2.378 ± 0.001$ hr (M. Hicks & T. Truong 2010),

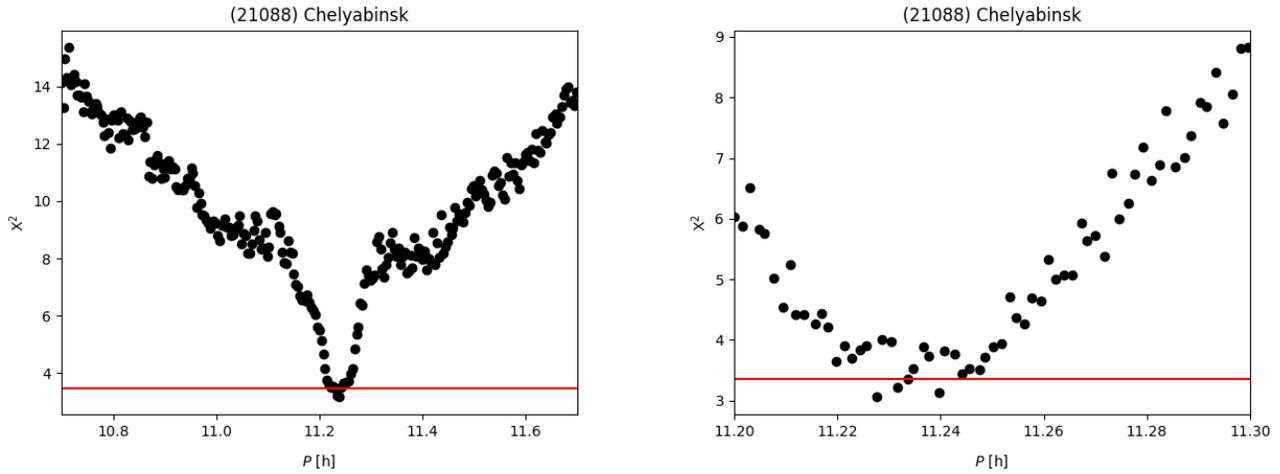

**Figure 6.** Output from the period search tool for (21088) Chelyabinsk. Each obtained period is plotted as a black circle, while a solid red line represents a 10% threshold from the lowest $\chi^2$ obtained in the search. Left: a wider interval from 10.7 to 11.7 hr is used to ensure the minimum of the data presented in this work. Right: the interval used is from 11.2 to 11.3 hr, with a coefficient $p = 0.8$.

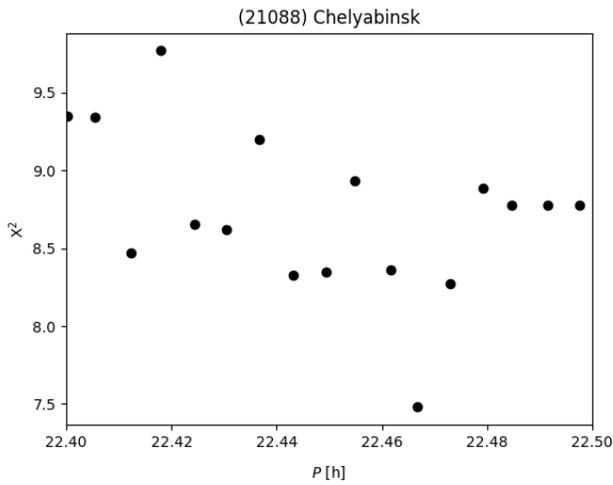

**Figure 7.** Output from the period search tool for (21088) Chelyabinsk. Each obtained period is plotted as a black circle. The interval used is from 22.4 to 22.5 hr, with a coefficient $p$ of 0.8.

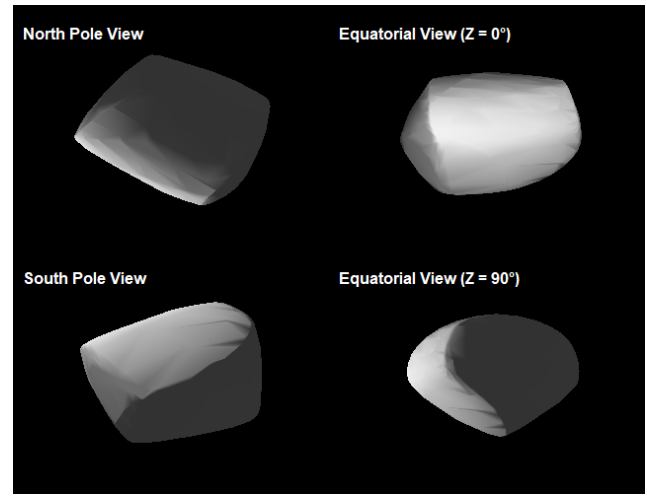

**Figure 8.** Obtained shape model of (21088) Chelyabinsk with a constant rotation period. Top left: north pole view ($Y$-axis = 0°). Bottom left: south pole view ($Y$-axis = 180°). Top right: equatorial view with $Z$-axis rotated 0°. Bottom right: equatorial view with $Z$-axis rotated 90°.

$P = 2.3774 \pm 0.0003$ hr (D. Higgins 2011), $P = 2.37760 \pm 0.00009$ hr (B. D. Warner 2018), $P = 2.37760 \pm 0.00009$ hr (B. D. Warner 2018), and $P = 2.3778 \pm 0.0003$ hr and $P = 2.3774 \pm 0.0003$ hr (J. Oey 2020). There are also web-suggested periods in the ALCDEF database that agree with the published ones, $P = 2.375 \pm 0.001$ hr (Higgins 2011web) and $P = 2.3775 \pm 0.0005$ hr (Behrend 2024web). We ran the tool several times, finding that the best solution is $P = 2.377471$ hr, as shown in Figure 10.

We next computed the medium search, adopting the obtained period as the starting one, obtaining the following medium solution: $P = 2.377471$ hr, $\lambda = 60°$, $\beta = -70°$, and $\chi^2_{\text{red}} = 1.42$ (in Appendix B.3 a graphical representation of the distribution of the obtained solutions is shown). With this medium solution, a fine solution was computed, obtaining $P = 2.377471$ hr, $\lambda = 64°$, $\beta = -71°$, $\chi^2_{\text{red}} = 1.41$, and $\epsilon \simeq 157°$, which implies a retrograde rotation. The best-fitting shape model with the data can be seen in Figure 11, and the graphical representation of the fit between the data and the model is shown in Figure 12 and Appendix C.3.

We finally computed the uncertainties. To do that, we created 100 different datasets from the initial dataset (3040 values), randomly removing 25% of the values, computing 100 models with them focusing on the medium solution. We obtained the following results: $P = 2.377471 \pm 0.000001$ hr $\lambda = 64° \pm 6°$, $\beta = -70° \pm 1°$, and $\epsilon = 157° \pm 1°$.

Since the time span is relatively large (14 yr), it was worth a try to check whether the asteroid is affected by YORP. For this, we followed the same method, maintaining the obtained period from the period scan tool ($P = 2.377471$ hr) since it has a good fit with the data. We used this time the code that allows the period to change linearly over time, where we obtained the following medium solution: $P = 2.377473$ hr, $\lambda = 60°$, $\beta = -70°$, $\upsilon = 1.75 \times 10^{-8}$ rad day$^{-2}$, and $\chi^2_{\text{red}} = 1.40$ (see Appendix B.4 for a graphical representation of the distribution of the solutions obtained). We computed a fine solution around this medium one, obtaining $P = 2.377473$ hr, $\lambda = 63°$, $\beta = -71°$, $\upsilon = 2.05 \times 10^{-8}$ rad day$^{-2}$, $\chi^2_{\text{red}} = 1.39$, and $\epsilon \simeq 157°$, which, again, implies a retrograde rotation. In

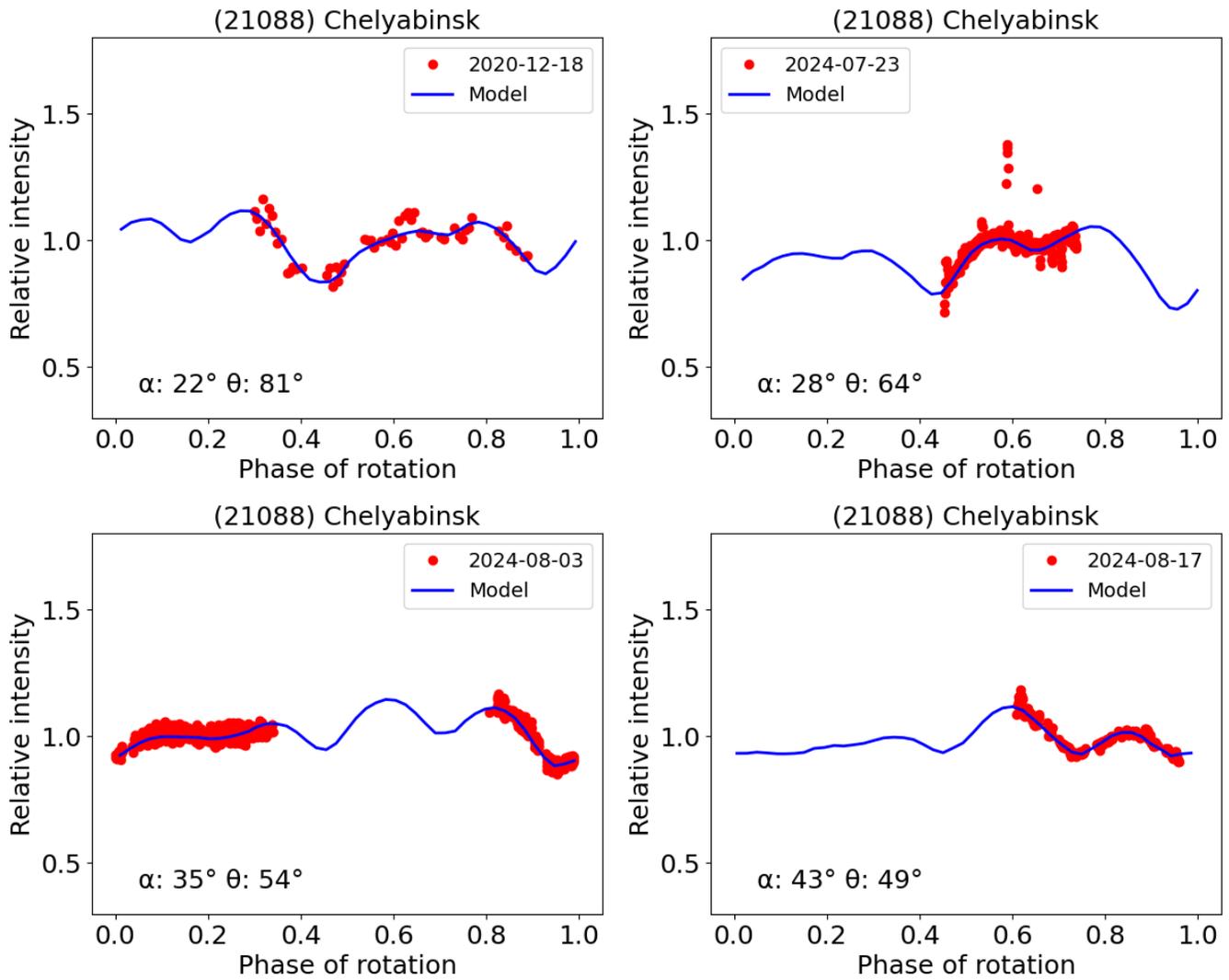

**Figure 9.** Graphical representation of the fit between four of the light curves used in this work (2020 December 18 and 2024 August 3 from ALCDEF and 2024 July 21 and August 17 from the newly obtained light curves) and the best-fitting constant-period model (C Model) for (21088) Chelyabinsk. The data are plotted as red circles for each observation, while the model is plotted as a solid blue line. The geometry is described by its solar phase angle $\alpha$ and its aspect angle $\theta$.

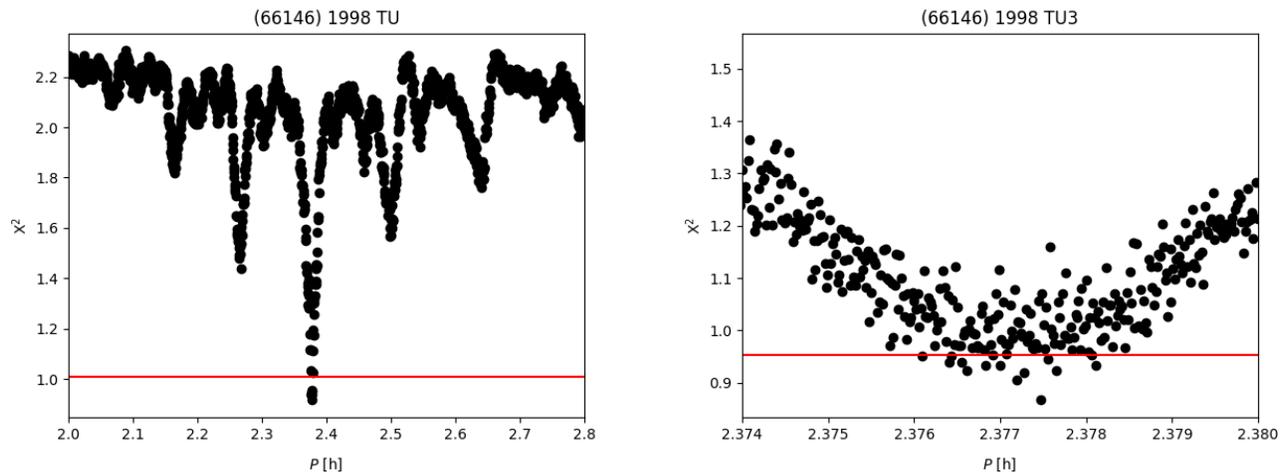

**Figure 10.** Output from the period search tool for (66146) 1998 TU3. Each obtained period is plotted as a black circle, while a solid red line represents a 10% threshold from the lowest $\chi^2$ obtained in the search. Left: a wider interval from 2 to 2.8 hr is used to ensure the minimum of the data presented in this work. Right: the interval used is from 2.37 to 2.38 hr, with a coefficient $p = 0.8$. There are several periods under the threshold, which implies that more data are needed to refine the value.

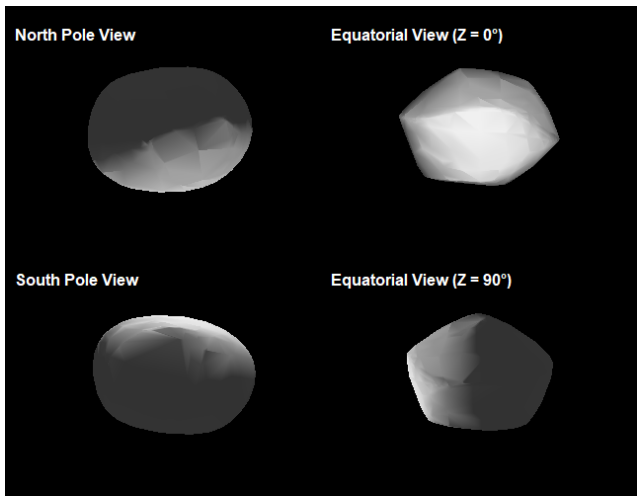

**Figure 11.** Obtained shape model of (66146) 1998 TU3 with a constant rotation period. Top left: north pole view ($Y$-axis = 0°). Bottom left: south pole view ($Y$-axis = 180°). Top right: equatorial view with $Z$-axis rotated 0°. Bottom right: equatorial view with $Z$-axis rotated 90°.

Figure 13, the best-fitting shape is shown, and in Figure 14 and Appendix C.4 a graphical representation of the fit between the data and the model is shown.

We proceeded in the same way as with the no-YORP model to compute the uncertainties, obtaining $P = 2.377473 \pm 0.000002$ hr $\lambda = 64° \pm 5°$, $\beta = -71° \pm 2°$, $\upsilon = (2.37 \pm 2.09) \times 10^{-8}$ rad day$^{-2}$, and $\epsilon = 157° \pm 1°$.

To support our claim that the YORP acceleration may be affecting (66146) 1998 TU3, we decided to iterate our best solution, as proposed in Section 3, with several values of $\upsilon$, from $-1 \times 10^{-8}$ to $5 \times 10^{-8}$ in $0.05 \times 10^{-8}$ steps; the results obtained are plotted in Figure 15. The values for $\upsilon$ obtained by this method ($\upsilon = (2.39 \pm 2.17) \times 10^{-8}$ rad day$^{-2}$) are in agreement with our solution of $\upsilon = (2.37 \pm 2.09) \times 10^{-8}$ rad day$^{-2}$.

Both models, constant period and linearly increasing period, have a good fit with the data, while the linearly increasing period model is slightly better, which makes us unable to rule out the possibility that the asteroid is affected by YORP, but more observations are needed to support this claim. As a way to rule out any of the two possibilities, the period was studied for different temporal spans of light curves (2010, 2017, 2019, and 2021), but the obtained period for each of them is a good enough match for both models, since the possible YORP effect detected is not strong, and the temporal span of ~14 yr is also short. Some clues suggesting that the YORP effect may be taking place could be the obliquity value of $\epsilon = 157°$ (not too far from the extreme value of 180°, a well-known consequence of the YORP effect taking place according to J. Hanuš et al. 2013), the proposed method to estimate the uncertainties of the YORP value, and the fact that all of the 100 models we created had a slightly better fit. Taking all that into account, we present both shape models and spin parameters. However, as this is a candidate rather than a confirmed YORP detection, we also note that this asteroid is worth studying in the future when more data are available.

Given that the inferred period change could be related to YORP, it is important to place its magnitude in the context of previously measured YORP accelerations. To do so, we compare (66146) 1998 TU3 with the sample of asteroids showing measured YORP accelerations compiled by J. Ďurech et al. (2024). With a diameter of ~3 km and a short rotation period of ~2.4 hr, 1998 TU3 lies near the upper size limit of objects for which YORP detections have been reported. According to theoretical expectations, YORP-driven angular accelerations decrease rapidly with increasing size, making detections for kilometer-scale asteroids challenging. The YORP acceleration derived here, $(2.4 \pm 2.1) \times 10^{-8}$ rad day$^{-2}$, is therefore small but physically plausible, and it is comparable in magnitude to those of other relatively large NEAs in the literature.

In addition, the two shape models obtained for (66146) 1998 TU3 exhibit noticeable morphological differences. The constant-period solution results in a more elongated body, whereas the model assuming a linearly increasing rotation period yields a slightly rounder shape with a more developed equatorial ridge. This behavior is physically consistent with the response of rapidly rotating asteroids, where centrifugal acceleration promotes mass redistribution toward the equatorial region. Similar morphologies have been observed in other fast-rotating NEAs, such as the well-known Bennu and Ryugu recently visited by OSIRIS-REx and the Hayabusa2 mission, respectively.

The existence of two viable solutions with comparable residuals suggests that both shapes capture the large-scale structure of the object at the resolution imposed by the light-curve data, while the inferred secular change in rotation period slightly modifies the inferred elongation. Despite its large size and rapid rotation, 1998 TU3 does not show photometric or morphological evidence for binarity or a contact binary configuration. Its moderate light-curve amplitude, together with the continuous equatorial ridge observed in both models, supports an interpretation as a single elongated body, potentially a rubble pile approaching rotational equilibrium rather than a binary system.

### 4.4. (297418) 2000 SP43

This is an asteroid that belongs to the PHA group, due to its MOID of 0.018918 au from data of the European Space Agency (ESA) Near Earth Objects Coordination Centre (NEOCC),[20] and it also belongs to the Atens group. There are four published diameters obtained using WISE data: $0.41 \pm 0.02$ km (C. R. Nugent et al. 2015), $0.407 \pm 0.019$ km (A. K. Mainzer et al. 2019), and $0.43 \pm 0.17$ km and $0.44 \pm 0.10$ km (J. R. Masiero et al. 2021).

There are five light curves available on the ALCDEF database from 2019 October 10 to 14, but there is no published model at the time of this work; hence, this would be the first shape model determination. Along with these archival data, we add eight more light curves in this paper, obtained in the framework of our ViNOS project, with a temporal span of 851 days, from 2020 June 19 to 2022 October 18. This makes a total of 13 light curves, with a temporal span of 3 yr.

First, we used those light curves to obtain a rotation period, taking into account the already-published ones: $P = 6.3136 \pm 0.0005$ hr (Pravec 2023web[21]), $P = 6.314 \pm 0.009$ hr (C. W. Hergenrother 2018), and $P = 6.306 \pm 0.002$ hr (B. D. Warner & R. D. Stephens 2020).

Taking into account these periods, we performed several searches around them, obtaining in most cases a minimum of $P = 6.313713$ hr, as Figure 16 shows. This is the value that is used as the initial period for all the subsequent models computed for this work.

---

[20] https://neo.ssa.esa.int/search-for-asteroids?sum=1&des=297418%202000SP43

[21] http://www.asu.cas.cz/~asteroid/297418.png

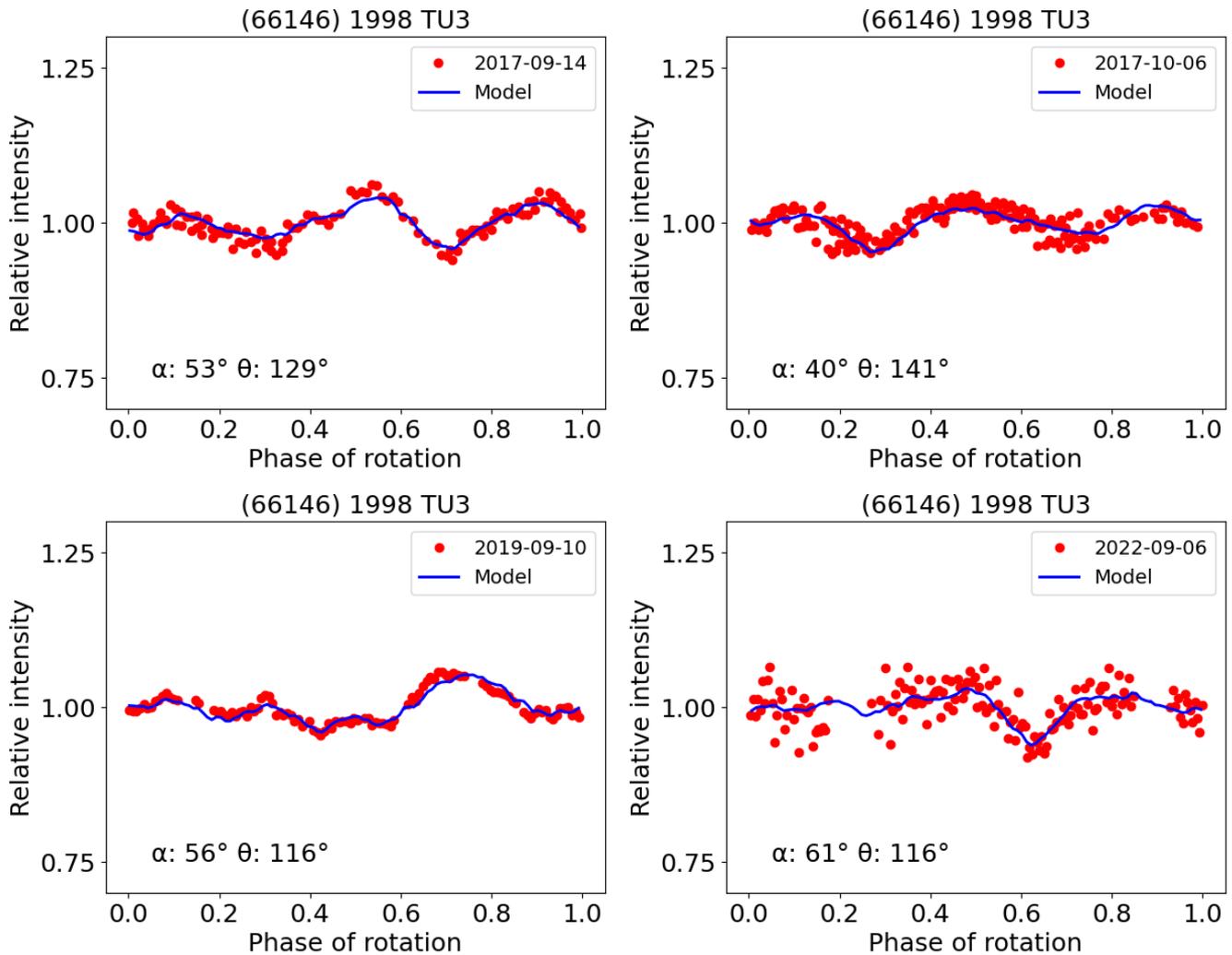

**Figure 12.** Graphical representation of the fit between four of the light curves used in this work (2017 September 14 and November 17 and 2019 September 20 from ALCDEF and 2022 September 6 from the newly obtained light curves) and the best-fitting constant-period (C Model) for (66146) 1998 TU3. The data are plotted as red circles for each observation, while the model is plotted as a solid blue line. The geometry is described by its solar phase angle $\alpha$ and its aspect angle $\theta$.

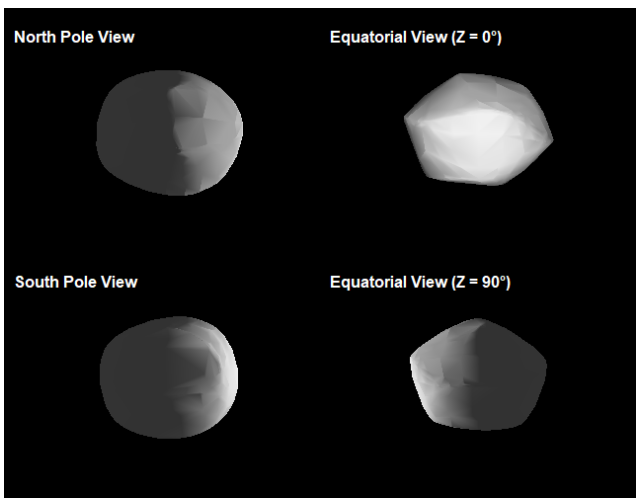

**Figure 13.** Obtained shape model of (66146) 1998 TU3 with a linearly increasing rotation period. Top left: north pole view ($Y$-axis = 0°). Bottom left: south pole view ($Y$-axis = 180°). Top right: equatorial view with $Z$-axis rotated 0°. Bottom right: equatorial view with $Z$-axis rotated 90°.

With this value, the medium search is computed, obtaining a primary solution of $P = 6.312231$ hr, $\lambda = 265°$, $\beta = 55°$, and $\chi^2_{\rm red} = 1.68$ (see Appendix B.5 for a graphical representation of the distribution of the solutions obtained). Taking these values as a starting point for the fine solution, we obtain $P = 6.312228$ hr, $\lambda = 265°$, $\beta = 54°$, $\chi^2_{\rm red} = 1.65$, and $\epsilon \simeq 26°$, which implies that this is a prograde rotator. The shape model with this fine solution is computed and presented in Figure 17, and the graphical representation of the fit between data and model is shown in Figure 18 and Appendix C.5. As the visual fit between this model and some of the figures seems to not be good enough, and since this result is the best we can obtain with the current data, we decided to remove those light curves and try to find a new model without them; nevertheless, we obtained almost the same values for all the parameters, thus making this model the best-fitting one in the whole set of data.

For the uncertainties for each value of the solution, we computed 100 different models. For this, we created 100 light-curve sets from the initial one (1596 values), randomly removing for each of them 25% of the values. We focused on the best medium solution, obtaining the following values: $P = 6.312233 \pm 0.000021$ hr, $\lambda = 265° \pm 5°$, $\beta = 54° \pm 4°$, and $\epsilon = 26° \pm 4°$. Since some of the light curves seem to not have a visual good fit to the model, we decided to, as for (4660) Nereus, search for secondary periods to conclude whether this asteroid is also a tumbler. In this case we were unable to find a good match with any secondary period; thus, we conclude that, with the current data, it is not possible to affirm that it is a tumbler. In an attempt to resolve whether those discrepancies are due to some features on the asteroids' surface that the inversion method cannot resolve, we decided to run a model without them, obtaining the same pole solution; thus, we conclude that the model here presented is the best-fitting one to the data.

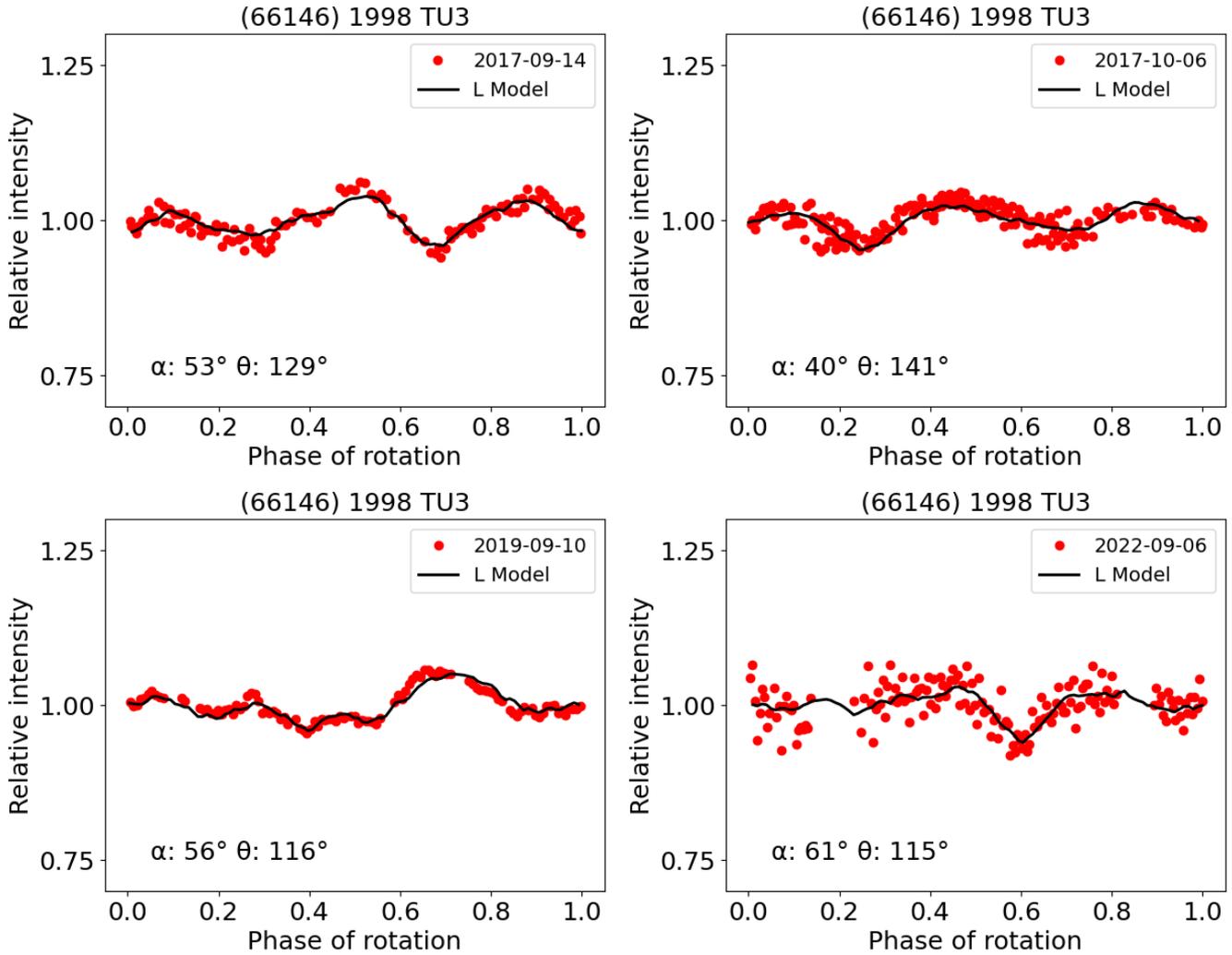

**Figure 14.** Graphical representation of the fit between four of the light curves used in this work (2017 September 14, 2017 October 6, and 2019 September 10 from ALCDEF and 2022 September 6 from the newly obtained light curves) and the best-fitting linearly increasing rotation period model (L Model) for (66146) 1998 TU3. The data are plotted as red circles for each observation, while the model is plotted as a solid black line. The geometry is described by its solar phase angle $\alpha$ and its aspect angle $\theta$.

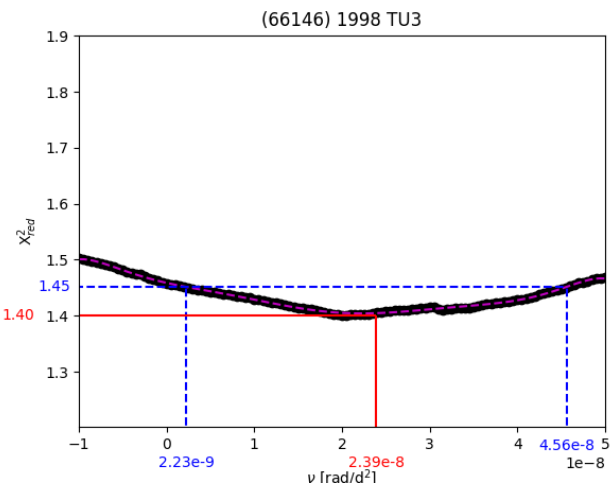

**Figure 15.** Representation of the change in $\chi^2_{\rm red}$ against $v$, in the range of $-1 \times 10^{-8}$ to $5 \times 10^{-8}$ for (66146) 1998 TU3, with the values of $\lambda$, $\beta$, and $P$ fixed in the best fine solution ($\lambda = 63°$, $\beta = -71°$, and $P = 2.377473$ hr). The best-fitting value (represented as red solid lines) is at $v = 2.39 \times 10^{-8}$ with a $\chi^2_{\rm red} = 1.40$. The $3\sigma$ value of $\chi^2_{\rm red} = 1.45$ corresponds to $v$ between $v = 0.22 \times 10^{-8}$ and $v = 4.56 \times 10^{-8}$ (represented as blue dashed lines).

The shape model of (297418) 2000 SP43 indicates an unusually elongated body. The large observed peak-to-peak light-curve amplitude ($\Delta m \sim 0.9$–$1.0$ mag) implies a minimum axial ratio of $a/b \approx 2.3 \pm 0.2$ (if assuming a triaxial ellipsoidal shape), placing this object among the most elongated NEAs with shape models derived from light-curve inversion. Such extreme elongation raises the possibility that 2000 SP43 could represent a contact binary or bilobed configuration similar to that of (8567) 1996 HW1 (C. Magri et al. 2011). While the convex inversion method employed here cannot directly resolve concavities, the combination of a very large light-curve amplitude and a highly elongated convex shape is consistent with this interpretation. However, with the currently available photometric data, it is not possible to unambiguously distinguish between a contact binary structure and a single, highly elongated body. Additional observations, such as radar imaging or dense light-curve coverage at multiple apparitions, would be required to further constrain its internal structure.

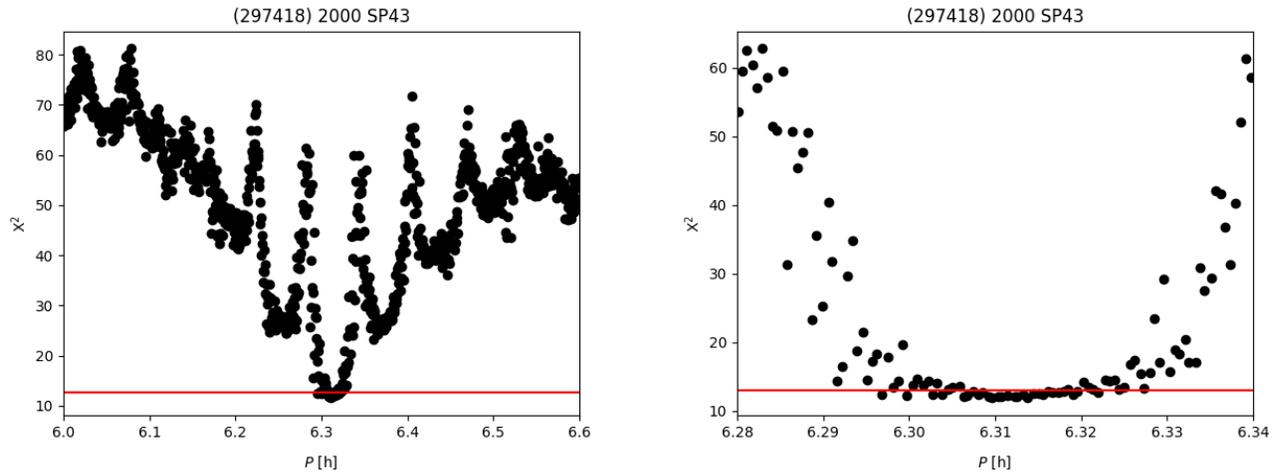

**Figure 16.** Output from the period search tool for (297418) 2000 SP43. Each obtained period is plotted as a black circle, while a solid red line represents a 10% threshold from the lowest $\chi^2$ obtained in the search. Left: a wider interval from 6 to 6.6 hr is used to ensure the minimum of the data presented in this work. Right: the interval used is from 6.28 to 6.34 hr, with a coefficient $p$ of 0.8. There are several periods under the threshold, which implies that more data are needed to refine the value.

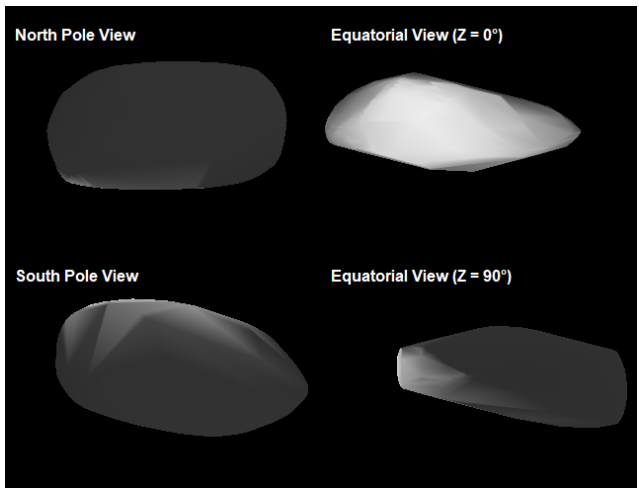

**Figure 17.** Obtained shape model of (297418) 2000 SP43 with a constant rotation period. Top left: north pole view ($Y$-axis = 0°). Bottom left: south pole view ($Y$-axis = 180°). Top right: equatorial view with $Z$-axis rotated 0°. Bottom right: equatorial view with $Z$-axis rotated 90°.

## 5. Results and Conclusions

In this work, we present 61 new light curves taken with three different telescopes at the Teide Observatory (Tenerife, Spain) and the Isaac Aznar/T35 telescopes (Valencia, Spain). Using these data and data obtained from the ALCDEF database (56 light curves; thus, ViNOS light curves are 52% of the curves used), we computed five new shape models, along with its spin parameters, one for (4660) Nereus, one for (21088) Chelyabinsk, two for (66146) 1998 TU3, and one for (297418) 2000 SP43. This is a significant contribution to the field, as only a few dozen NEA shape models are currently published (P. Fatka et al. 2025).

For (4660) Nereus we found a new pole solution with $P = 15.159442 \pm 0.000738$ hr, $\lambda = 321° \pm 19°$, $\beta = 78° \pm 12°$, and $\epsilon = 12° \pm 11°$. The period agrees with the previously published ones, while the poles ($\lambda$, $\beta$) differ by a margin from the published radar model from M. Brozovic et al. (2009); those values are still within the possible values for the data presented in this work. Our shape model, shown in Figure 4, is an overall good match to the one presented in that work, presenting similarities in the elongated shape in the $X$-axis and the same features in the top and bottom views ($Z$-axis). Finally, due to some discrepancies between the data and the model, we performed a search for secondary periods in the residual of the model, obtaining a secondary period of $12{,}4032 \pm 0{,}0011$, which is in agreement with previous publications (B. D. Warner & R. D. Stephens 2022).

For (21088) Chelyabinsk we found $P = 11.227651 \pm 0.000025$ hr, $\lambda = 232° \pm 2°$, $\beta = -55° \pm 1°$, and $\epsilon = 108° \pm 1°$. The period is in agreement with the already-published ones, except those found by Pravec (Pravec 2004web) of $P = 22.431 \pm 0.001$ hr and $P = 22.49$ hr. We tried several searches around this value, but the fit was always poorer than our presented solution of $P = 11.227651 \pm 0.000025$ hr as shown in Figures 6 and 7.

For (66146) 1998 TU3 we derived two models, one with constant period ($P = 2.377471 \pm 0.000001$ hr, $\lambda = 64° \pm 6°$, $\beta = -70° \pm 1°$, and $\epsilon = 157° \pm 1°$) and one with YORP acceleration taking effect ($P = 2.377473 \pm 0.000002$ hr, $\lambda = 64° \pm 5°$, $\beta = -71° \pm 2°$, $\upsilon = (2.37 \pm 2.09) \times 10^{-8}$ rad day$^{-2}$, and $\epsilon = 157° \pm 1°$). Their fit to the data is similar, while the model taking YORP into account fits the data slightly better. Our models consistently find that the YORP acceleration is present. This would add (66146) 1998 TU3 to the short list of asteroids under the YORP effect: (1862) Apollo (M. Kaasalainen et al. 2007), (54509) 2000 PH5

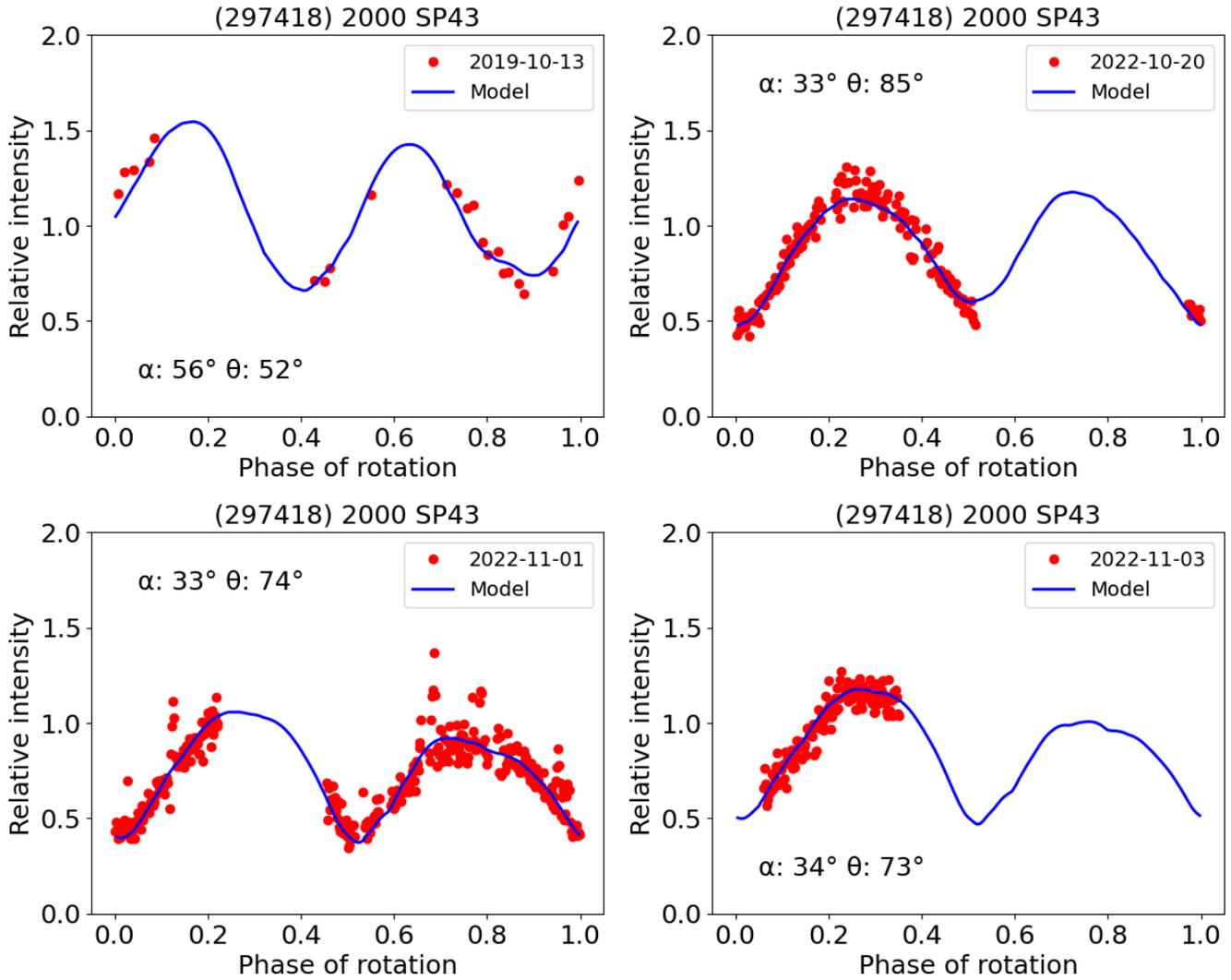

**Figure 18.** Graphical representation of the fit between four of the light curves used in this work (2019 October 13 from ALCDEF and 2022 October 20, November 1, and November 3 from the newly obtained light curves) and the best-fitting constant-period model (C Model) for (297418) 2000 SP43. The data are plotted as red circles for each observation, while the model is plotted as a solid blue line. The geometry is described by its solar phase angle $\alpha$ and its aspect angle $\theta$.

(S. C. Lowry et al. 2007; P. A. Taylor et al. 2007), (1620) Geographos (J. Ďurech et al. 2008b), (25143) Itokawa (S. C. Lowry et al. 2014), (1685) Toro, (3103) Eger, and (161989) Cacus (J. Ďurech et al. 2018), as well as the candidates (2100) Ra-Shalom (J. Rodríguez Rodríguez et al. 2024a; J. Ďurech et al. 2024) and (159402) 1999 AP10 (J. Rodríguez Rodríguez et al. 2024b), with (159402) 1999 AP10 being to this date the only one known to be decelerated rather than accelerated like the others, including (66146) 1998 TU3. It is also worth mentioning that while for these data these solutions have the best fit, we cannot rule out the possibility of another solution with $\lambda \sim 270$, as shown in Figures 22 and 21, due to the spherical shape this asteroid presents, as shown in Figures 13 and 11. We notice also that both solutions give a rounder shape with a developed equatorial ridge, a behavior physically consistent with the response of rapidly rotating asteroids, where centrifugal acceleration promotes mass redistribution toward the equatorial region. Its moderate light-curve amplitude, together with the shape models, supports an interpretation as a single elongated body, potentially a rubble pile approaching rotational equilibrium.

For (297418) 2000 SP43 the best-fitting solution we found is $P = 6.312233 \pm 0.000021$ hr, $\lambda = 265° \pm 5°$, $\beta = 54° \pm 4°$, and $\epsilon = 26° \pm 4°$. Since some of the light curves seem to not have a visual good fit to the model, we made an attempt to find any secondary period in the residuals of the model, without success. Finally, we created another model without those light curves, obtaining the same pole solution as using all the data. The observed very large amplitude of the light curves ($\Delta m \sim 0.9$–1.0 mag), which implies a minimum axial ratio of $a/b \approx 2.3 \pm 0.2$, and the very elongated shape model obtained suggest that (297418) 2000 SP43 could be a contact binary, but more data are needed to confirm this hypothesis.

Last but not least, the spin pole orientations derived for (4660) Nereus and (297418) 2000 SP43 are consistent with the sense of rotation inferred from measured Yarkovsky accelerations listed in the Small-Body Database, providing an independent validation of the pole solutions.

## Acknowledgments

We thank Dr. Josef Ďurech for providing us with the inversion code that includes the Yarkovsky–O'Keefe–

Radzievskii–Paddack (YORP) acceleration and for his advice in using the inversion codes. We also thank the anonymous referee for the useful comments that helped to improve this manuscript.

J.L., M.R.A., and M.S.-R. acknowledge support from the Agencia Estatal de Investigacion del Ministerio de Ciencia e Innovacion (AEI-MCINN) under grant "Hydrated Minerals and Organic Compounds in Primitive Asteroids" with reference PID2020-120464GB-100.

This article includes observations made with the Two-meter Twin Telescope (TTT[22]) sited at the Teide Observatory of the Instituto de Astrofísica de Canarias (IAC), which Light Bridges operates in Tenerife, Canary Islands (Spain). The Observing Time Rights (DTO) used for this research were provided by RICTEL TTT, SA. M.S.-R. used storage and computing capacity in ASTRO POC's EDGE computing center at Tenerife under the form of Indefeasible Computer Rights (ICR). The ICR were provided by Light Bridges, SL with the collaboration of Hewlett Packard Enterprise and VAST DATA. This article includes observations made in the IAC80 telescope sited at the Teide Observatory of the Instituto de Astrofísica de Canarias (IAC), Canary Islands (Spain), and operated by the IAC.

This research was supported by Hunosa through the SPECIFIC COLLABORATION AGREEMENT FOR THE PROMOTION OF RESEARCH ON SPACE MINING AND ENERGY RESOURCES, reference SV-21-HUNOSA-2.

This work uses data obtained from the Asteroid Lightcurve Data Exchange Format (ALCDEF) database, which is supported by funding from NASA grant 80NSSC18K0851.

This work uses software MPO LC invert from Brian Warner to plot the asteroid shapes as presented in Figures 2, 8, 11, 13, and 17.

*Facilities:* HST (STIS), Swift (XRT and UVOT), AAVSO, CTIO:1.3m, CTIO:1.5m, CXO.

*Software:* Astroquery (A. Ginsburg et al. 2019), PyAstronomy (S. Czesla et al. 2019), DAMIT (J. Ďurech et al. 2010).

## Author Contributions

All authors contributed equally to this collaboration.

---

## Appendix A
## Summary LC

In this appendix, we share the tables with the ephemerides for each observation obtained from the ALCDEF database (Tables 4–7). In each table and for each observation, the observational circumstances are shown with their respective references.

### A.1. (4660) Nereus Archive Light Curves

**Table 4**
Archival Data Obtained from ALCDEF for Object (4660) Nereus

| Date | UT (Start) | UT (End) | $\alpha$ (deg) | $r$ (au) | $\Delta$ (au) | PABLon (deg) | PABLat (deg) | Reference |
|---|---|---|---|---|---|---|---|---|
| 1997 Aug 2 | 23:29:55.450 | 8:18:17.510 | 8.47 | 2.0110 | 1.0185 | 297.97 | −0.46 | ISH00 |
| 1997 Aug 4 | 00:28:08.256 | 8:34:14.131 | 9.15 | 2.0100 | 1.0215 | 297.92 | −0.45 | ISH00 |
| 2021 Oct 26 | 06:39:46.800 | 1:54:15.264 | 15.98 | 1.1967 | 0.2125 | 43.10 | 4.75 | WAR22 |
| 2021 Oct 27 | 03:59:57.523 | 0:48:55.814 | 15.56 | 1.1920 | 0.2073 | 43.52 | 4.84 | WAR22 |
| 2021 Oct 28 | 20:23:08.419 | 4:52:11.424 | 14.79 | 1.1832 | 0.1975 | 44.34 | 5.02 | WAR22 |
| 2021 Oct 29 | 03:57:36.605 | 2:42:33.206 | 14.64 | 1.1815 | 0.1957 | 44.48 | 5.06 | WAR22 |
| 2021 Oct 30 | 03:41:17.434 | 2:50:57.869 | 14.21 | 1.1763 | 0.1901 | 44.96 | 5.17 | WAR22 |
| 2021 Nov 3 | 03:31:10.042 | 0:49:45.408 | 12.69 | 1.1557 | 0.1684 | 46.94 | 5.67 | WAR22 |

**Note.** The table includes, in the following order, the date, starting time and end time (UT) of the observations, phase angle ($\alpha$), heliocentric ($r$) and geocentric ($\Delta$) distances, and phase-angle bisector longitude (PABLon) and latitude (PABLat) of the asteroid at the time of observation. References: ISH00: Y. Ishibashi et al. (2000); WAR22: B. D. Warner & R. D. Stephens (2022).

### A.2. (21088) Chelyabinsk Archive Light Curves

**Table 5**
Archival Data Obtained from ALCDEF for Object (21088) Chelyabinsk

| Date | UT (Start) | UT (End) | $\alpha$ (deg) | $r$ (au) | $\Delta$ (au) | PABLon (deg) | PABLat (deg) | Reference |
|---|---|---|---|---|---|---|---|---|
| 2020 Dec 15 | 08:06:17.885 | 2:30:02.650 | 22.82 | 2.0087 | 1.2498 | 124.14 | 5.35 | WAR21 |
| 2020 Dec 16 | 07:05:30.941 | 3:31:46.358 | 22.42 | 2.0104 | 1.2412 | 124.17 | 5.12 | WAR21 |
| 2020 Dec 18 | 06:49:59.030 | 3:26:10.090 | 21.54 | 2.0139 | 1.2239 | 124.22 | 4.65 | WAR21 |
| 2020 Dec 19 | 06:34:44.832 | 3:32:42.259 | 21.09 | 2.0156 | 1.2156 | 124.23 | 4.41 | WAR21 |
| 2024 Aug 1 | 21:37:07.565 | 0:07:30.403 | 33.52 | 1.3026 | 0.3699 | 316.70 | 27.58 | FAR25 |
| 2024 Aug 2 | 21:36:55.210 | 2:59:13.661 | 34.23 | 1.3021 | 0.3742 | 316.73 | 28.51 | BOT24 |
| 2024 Aug 3 | 22:00:24.394 | 4:00:11.952 | 34.96 | 1.3017 | 0.3788 | 316.75 | 29.44 | GOM24 |

**Notes.** The table includes, in the following order, the date, starting time and end time (UT) of the observations, phase angle ($\alpha$), heliocentric ($r$) and geocentric ($\Delta$) distances, and phase-angle bisector longitude (PABLon) and latitude (PABLat) of the asteroid at the time of observation. *Data available on ALCDEF with no manuscript associated.
**References:** WAR21:B. D. Warner & R. D. Stephens (2021); BOT24*; GOM24*; FAR25: R. G. Farfán et al. (2025).

*A.3. (66146) 1998 TU3 Archive Light Curves*

**Table 6**
Archival Data Obtained from ALCDEF for Object (66146) 1998 TU3

| Date | UT (Start) | UT (End) | α (deg) | r (au) | Δ (au) | PABLon (deg) | PABLat (deg) | Reference |
|------|-----------|----------|---------|--------|--------|--------------|--------------|-----------|
| 2008 Oct 22 | 11:21:38.477 | 2:21:36.518 | 88.97 | 0.9379 | 0.3497 | 93.84 | −11.21 | SKI12 |
| 2010 Sep 18 | 14:21:08.582 | 9:15:32.717 | 45.97 | 1.1672 | 0.2586 | 25.29 | −16.29 | HIG11 |
| 2010 Sep 20 | 15:45:39.053 | 9:06:53.712 | 44.15 | 1.1678 | 0.2491 | 25.14 | −16.72 | HIG11 |
| 2010 Sep 23 | 12:49:49.958 | 6:00:59.818 | 41.52 | 1.1679 | 0.2362 | 24.75 | −17.38 | HIG11 |
| 2010 Sep 23 | 16:21:39.658 | 8:22:10.733 | 41.39 | 1.1679 | 0.2356 | 24.73 | −17.42 | HIG11 |
| 2010 Sep 24 | 12:43:05.952 | 4:55:33.283 | 40.60 | 1.1678 | 0.2320 | 24.57 | −17.62 | HIG11 |
| 2010 Sep 27 | 13:30:42.509 | 7:25:35.558 | 37.80 | 1.1667 | 0.2196 | 23.83 | −18.33 | HIG11 |
| 2010 Sep 27 | 17:29:22.790 | 9:11:48.509 | 37.66 | 1.1666 | 0.2190 | 23.79 | −18.37 | HIG11 |
| 2017 Sep 5 | 08:52:31.238 | 1:07:53.357 | 57.41 | 1.1618 | 0.3843 | 28.94 | −12.24 | WAR18 |
| 2017 Sep 12 | 11:20:01.795 | 2:28:25.190 | 54.25 | 1.1672 | 0.3422 | 31.05 | −13.40 | WAR18 |
| 2017 Sep 14 | 08:35:07.872 | 2:29:06.144 | 53.35 | 1.1677 | 0.3308 | 31.54 | −13.75 | WAR18 |
| 2017 Sep 15 | 08:29:55.622 | 2:30:19.238 | 52.85 | 1.1679 | 0.3248 | 31.78 | −13.93 | WAR18 |
| 2017 Sep 16 | 08:27:35.914 | 2:30:03.946 | 52.35 | 1.1679 | 0.3188 | 32.01 | −14.13 | WAR18 |
| 2017 Sep 17 | 08:41:09.542 | 2:33:00.029 | 51.82 | 1.1678 | 0.3127 | 32.24 | −14.33 | WAR18 |
| 2017 Sep 18 | 08:22:48.461 | 2:32:31.344 | 51.30 | 1.1677 | 0.3067 | 32.44 | −14.53 | WAR18 |
| 2017 Oct 2 | 15:30:21.139 | 8:49:47.280 | 42.61 | 1.1537 | 0.2234 | 33.69 | −18.14 | OEY20 |
| 2017 Oct 5 | 16:44:25.901 | 8:44:03.062 | 40.65 | 1.1479 | 0.2072 | 33.36 | −19.11 | OEY20 |
| 2017 Oct 6 | 11:48:19.469 | 7:59:19.997 | 40.15 | 1.1462 | 0.2031 | 33.22 | −19.37 | OEY20 |
| 2017 Oct 7 | 11:54:39.802 | 7:28:31.382 | 39.54 | 1.1440 | 0.1980 | 33.02 | −19.71 | OEY20 |
| 2019 Aug 31 | 17:19:08.314 | 8:24:55.238 | 73.59 | 1.0307 | 0.0876 | 318.08 | −37.39 | OEY20 |
| 2019 Aug 31 | 18:21:31.162 | 9:16:19.546 | 73.46 | 1.0310 | 0.0878 | 318.09 | −37.30 | OEY20 |
| 2019 Sep 3 | 18:02:19.018 | 9:23:31.027 | 65.31 | 1.0469 | 0.1014 | 318.80 | −31.74 | OEY20 |
| 2019 Sep 4 | 18:05:05.597 | 8:41:22.963 | 63.24 | 1.0520 | 0.1066 | 319.13 | −30.21 | OEY20 |
| 2019 Sep 8 | 08:53:46.666 | 9:55:04.973 | 57.88 | 1.0695 | 0.1274 | 320.49 | −25.77 | OEY20 |
| 2019 Sep 10 | 08:51:00.691 | 1:09:10.771 | 55.99 | 1.0785 | 0.1398 | 321.30 | −23.90 | OEY20 |
| 2019 Sep 10 | 11:25:23.808 | 2:35:44.362 | 55.90 | 1.0790 | 0.1404 | 321.35 | −23.80 | OEY20 |
| 2019 Sep 10 | 16:28:03.274 | 8:20:58.070 | 55.74 | 1.0799 | 0.1418 | 321.44 | −23.64 | OEY20 |
| 2019 Sep 11 | 09:48:00.144 | 1:07:57.677 | 55.23 | 1.0831 | 0.1464 | 321.73 | −23.04 | OEY20 |
| 2019 Sep 11 | 09:48:00.144 | 3:28:59.866 | 55.23 | 1.0831 | 0.1464 | 321.73 | −23.04 | OEY20 |
| 2019 Sep 11 | 11:43:21.216 | 4:11:05.856 | 55.17 | 1.0834 | 0.1469 | 321.77 | −22.98 | OEY20 |
| 2019 Sep 11 | 15:30:56.563 | 8:07:51.312 | 55.07 | 1.0841 | 0.1479 | 321.84 | −22.87 | OEY20 |
| 2019 Sep 24 | 09:21:59.242 | 0:50:21.005 | 52.47 | 1.1295 | 0.2361 | 327.19 | −16.51 | OEY20 |
| 2019 Sep 25 | 09:30:21.571 | 0:43:28.358 | 52.52 | 1.1323 | 0.2434 | 327.61 | −16.20 | OEY20 |
| 2019 Sep 26 | 09:05:39.984 | 0:42:47.750 | 52.59 | 1.1350 | 0.2505 | 328.03 | −15.90 | OEY20 |
| 2019 Sep 27 | 09:24:42.365 | 5:44:21.379 | 52.68 | 1.1376 | 0.2579 | 328.46 | −15.61 | OEY20 |
| 2019 Oct 31 | 09:56:38.285 | 0:18:57.485 | 58.03 | 1.1628 | 0.5032 | 343.60 | −10.27 | OEY20 |

**Note.** The table includes, in the following order, the date, starting time and end time (UT) of the observations, phase angle (α), heliocentric (r) and geocentric (Δ) distances, and phase-angle bisector longitude (PABLon) and latitude (PABLat) of the asteroid at the time of observation.
**References.** SKI12: B. A. Skiff et al. (2012); HIG11: D. Higgins (2011); WAR18: B. D. Warner (2018); OEY20: J. Oey (2020).

*A.4. (297418) 2000 SP43 Archive Light Curves*

**Table 7**
Archival Data Obtained from ALCDEF for Object (297418) 2000 SP43

| Date | UT (Start) | UT (End) | α (deg) | r (au) | Δ (au) | PABLon (deg) | PABLat (deg) | Reference |
|------|-----------|----------|---------|--------|--------|--------------|--------------|-----------|
| 2019 Oct 10 | 02:19:34.147 | 6:17:05.309 | 57.78 | 1.0859 | 0.1872 | 339.87 | 11.99 | WAR20 |
| 2019 Oct 11 | 02:24:31.968 | 6:34:11.914 | 57.28 | 1.0905 | 0.1954 | 340.76 | 12.00 | WAR20 |
| 2019 Oct 12 | 02:25:00.826 | 6:21:36.259 | 56.84 | 1.0949 | 0.2036 | 341.61 | 12.02 | WAR20 |
| 2019 Oct 13 | 02:28:01.315 | 6:36:46.915 | 56.45 | 1.0993 | 0.2119 | 342.42 | 12.03 | WAR20 |
| 2019 Oct 14 | 02:34:35.904 | 6:35:06.950 | 56.12 | 1.1035 | 0.2203 | 343.21 | 12.04 | WAR20 |

**Note.** The table includes, in the following order, the date, starting time and end time (UT) of the observations, phase angle (α), heliocentric (r) and geocentric (Δ) distances, and phase-angle bisector longitude (PABLon) and latitude (PABLat) of the asteroid at the time of observation.
**References.** WAR20: B. D. Warner & R. D. Stephens (2020)

## Appendix B
## Pole Plots

In this appendix, we share the pole plot for each asteroid, showing the fit (in terms of $\chi^2_{red}$) between each pair of ($\lambda$, $\beta$) and the data (Figures 19–23). The red highlighted poles are within a margin of 10% from the lowest $\chi^2_{red}$ for each asteroid.

*B.1. (4660) Nereus Pole Search*

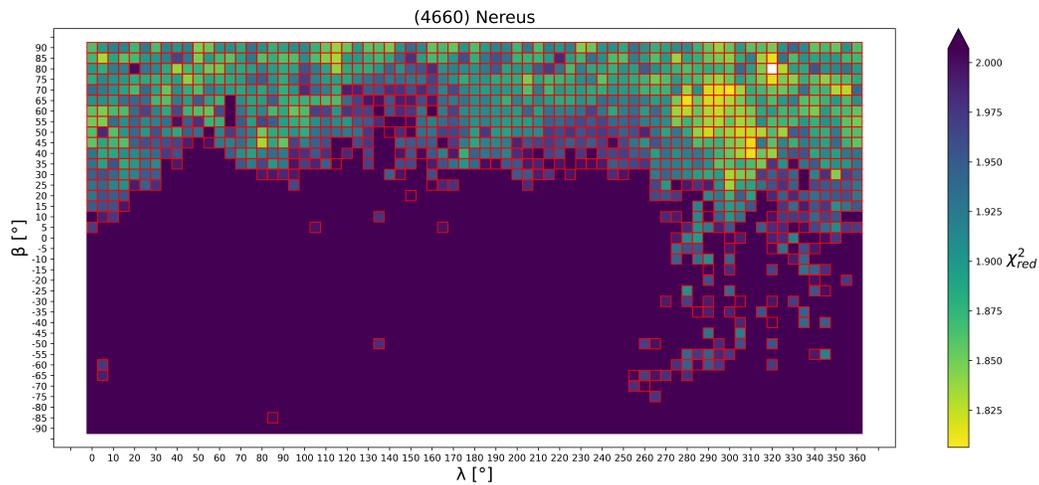

**Figure 19.** The statistical quality of the pole solutions for (4660) Nereus applying the constant-period code. The solutions are shaded by its $\chi^2_{red}$ value, with the best solution obtained represented as a white square ($\lambda = 290°$, $\beta = 70°$) with $\chi^2_{red} = 1.48$ (normalized for the 2500 data points). The solutions highlighted with a red border are within a margin of 10% ($3\sigma$) of the best obtained solution.

*B.2. (21088) Chelyabinsk Pole Search*

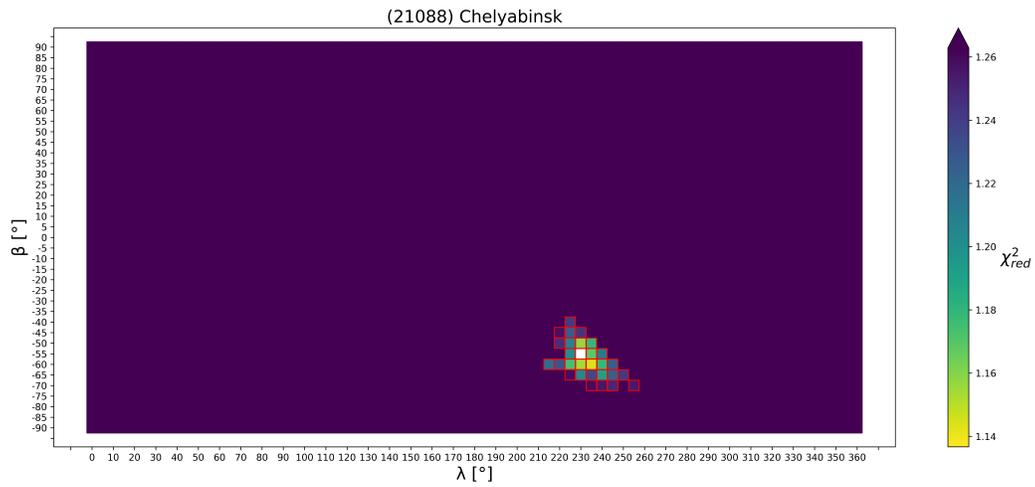

**Figure 20.** The statistical quality of the pole solutions for (21088) Chelyabinsk applying the constant-period code. The solutions are shaded by its $\chi_{red}^2$ value, with the best solution obtained represented as a white square ($\lambda = 230°$, $\beta = -55°$) with $\chi_{red}^2 = 1.14$ (normalized for the 3040 data points). The solutions highlighted with a red border are within a margin of 10% ($3\sigma$) of the best obtained solution.

*B.3. (66146) 1998 TU3 Pole Search (Constant Period)*

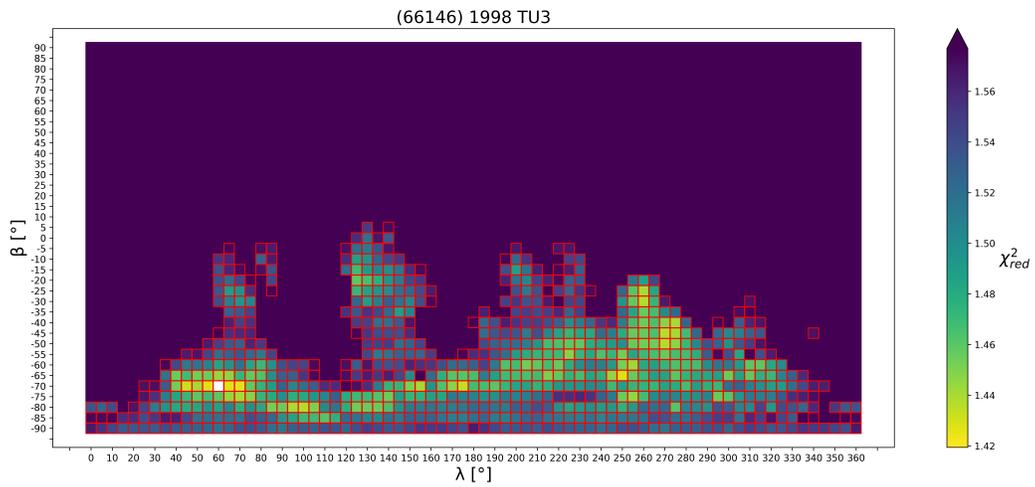

**Figure 21.** The statistical quality of the pole solutions for (66146) 1998 TU3 applying the constant-period code. The solutions are shaded by its $\chi_{red}^2$ value, with the best solution obtained represented as a white square ($\lambda = 60°$, $\beta = -70°$) with $\chi_{red}^2 = 1.42$ (normalized for the 3040 data points). The solutions highlighted with a red border are within a margin of 10% ($3\sigma$) of the best obtained solution.

### B.4. (66146) 1998 TU3 Pole Search (Linearly Increasing Period)

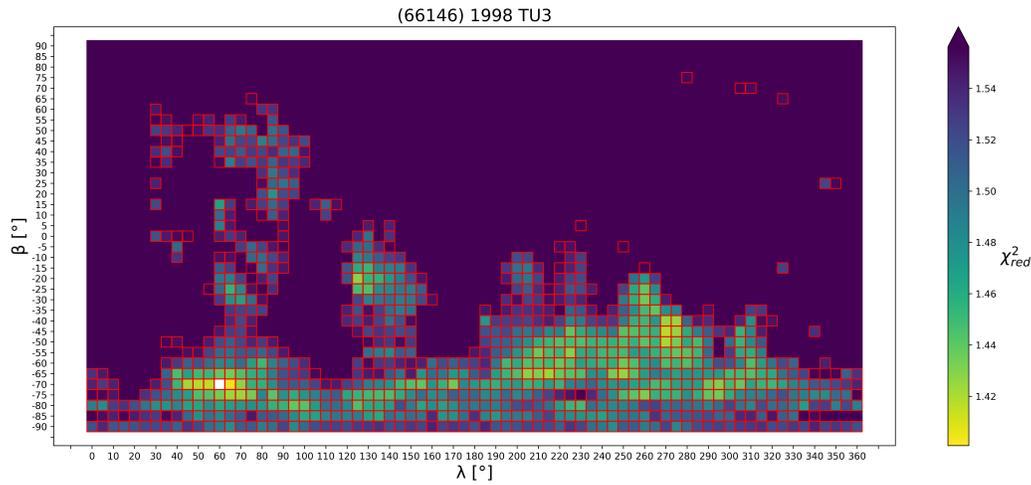

**Figure 22.** The statistical quality of the pole solutions for (66146) 1998 TU3 applying the linearly increasing period code. The solutions are shaded by its $\chi^2_{red}$ value, with the best solution obtained represented as a white square ($\lambda = 60°$, $\beta = -70°$) with $\chi^2_{red} = 1.40$ (normalized for the 3040 data points). The solutions highlighted with a red border are within a margin of 10% ($3\sigma$) of the best obtained solution.

### B.5. (297418) 2000 SP43 Pole Search

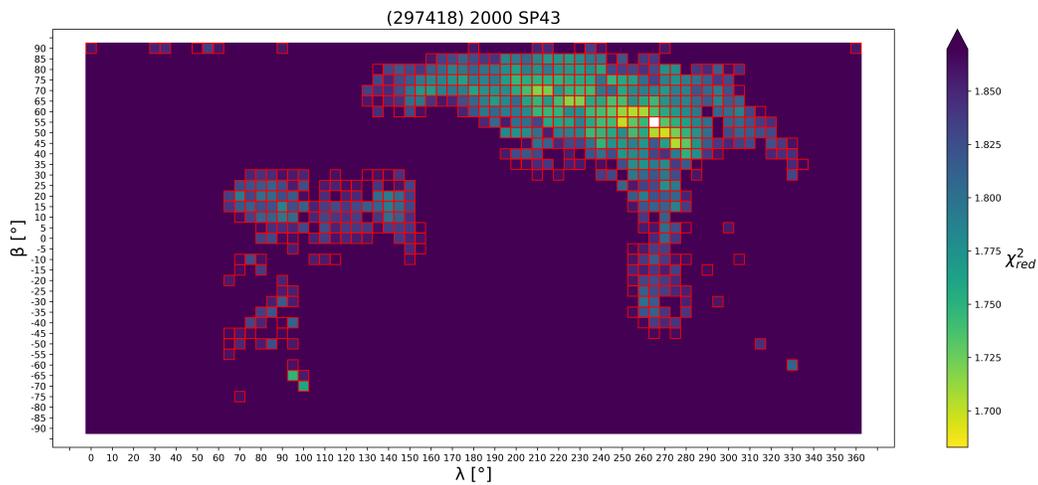

**Figure 23.** The statistical quality of the pole solutions for (297418) 2000 SP43 applying the constant-period code. The solutions are shaded by its $\chi^2_{red}$ value, with the best solution obtained represented as a white square ($\lambda = 265°$, $\beta = 55°$) with $\chi^2_{red} = 1.68$ (normalized for the 1596 data points). The solutions highlighted with a red border are within a margin of 10% ($3\sigma$) of the best obtained solution.

## Appendix C
## Fits

In this appendix, we share a graphical representation of the data and the synthetic light curves created from the best model (Figures 24–28).

### C.1. (4660) Nereus Model Fit

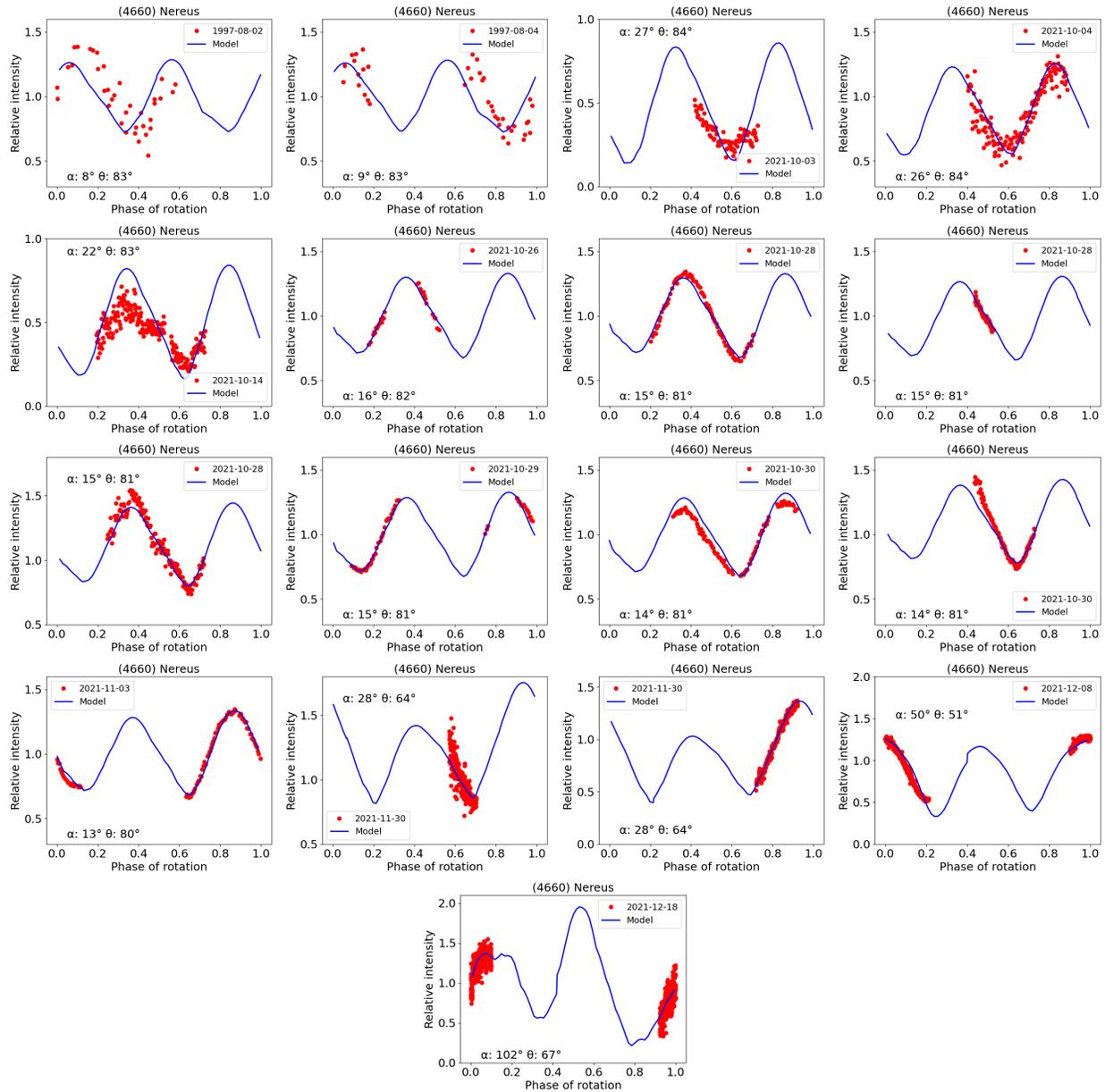

**Figure 24.** Fit between the rest (15 light curves) of the light curves from (4660) Nereus used for creating the shape model in this work and the best-fitting constant-period model (C Model). The data are plotted as red circles for each observation, while the model is plotted as a solid blue line. The geometry is described by its solar phase angle $\alpha$ and its aspect angle $\theta$.

## C.2. (21088) Chelyabinsk Model Fit

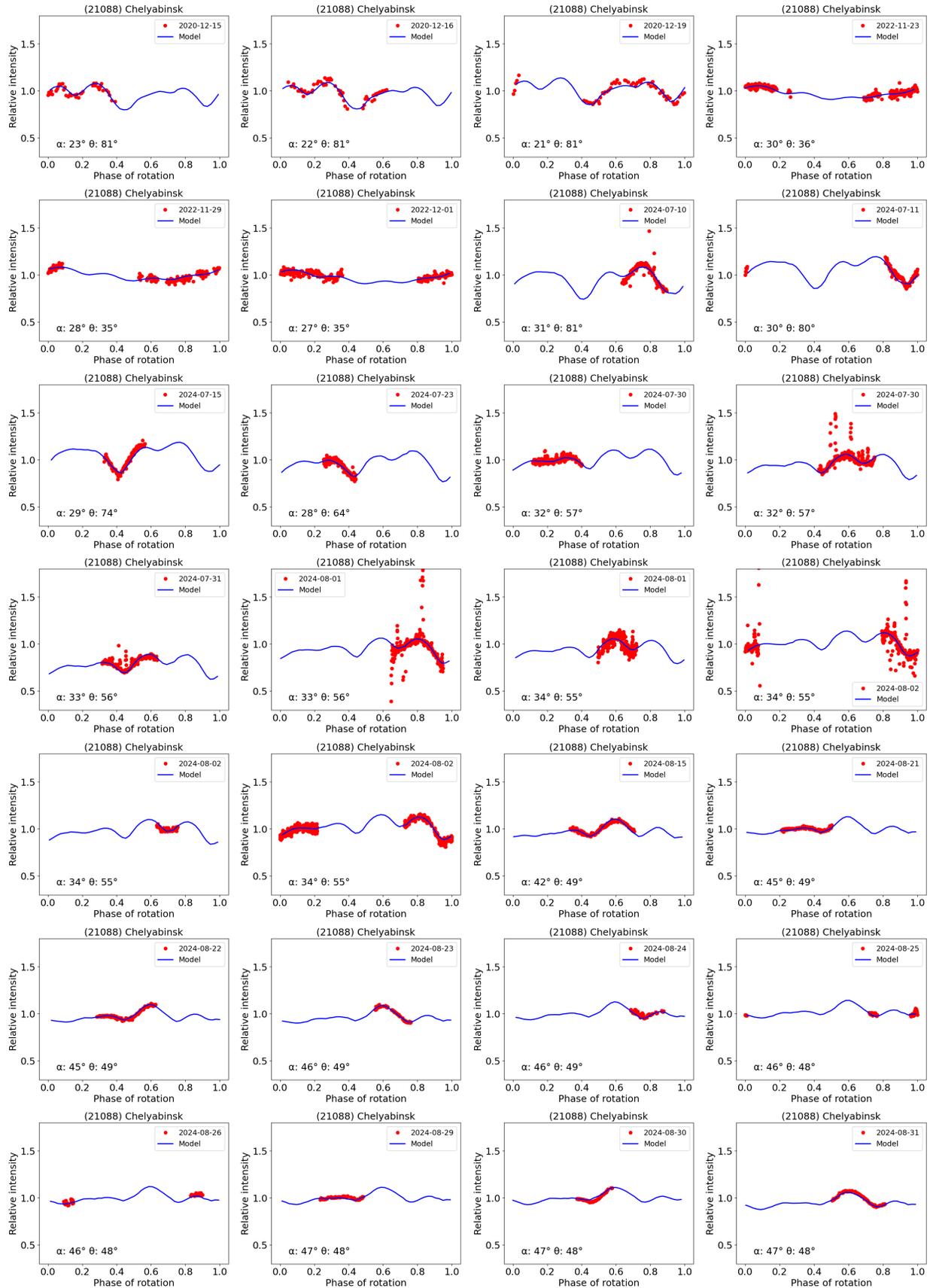

**Figure 25.** Fit between the rest (41 light curves) of the light curves from (21088) Chelyabinsk used for creating the shape model in this work and the best-fitting constant-period model (C Model). The data are plotted as red circles for each observation, while the model is plotted as a solid blue line. The geometry is described by its solar phase angle $\alpha$ and its aspect angle $\theta$.

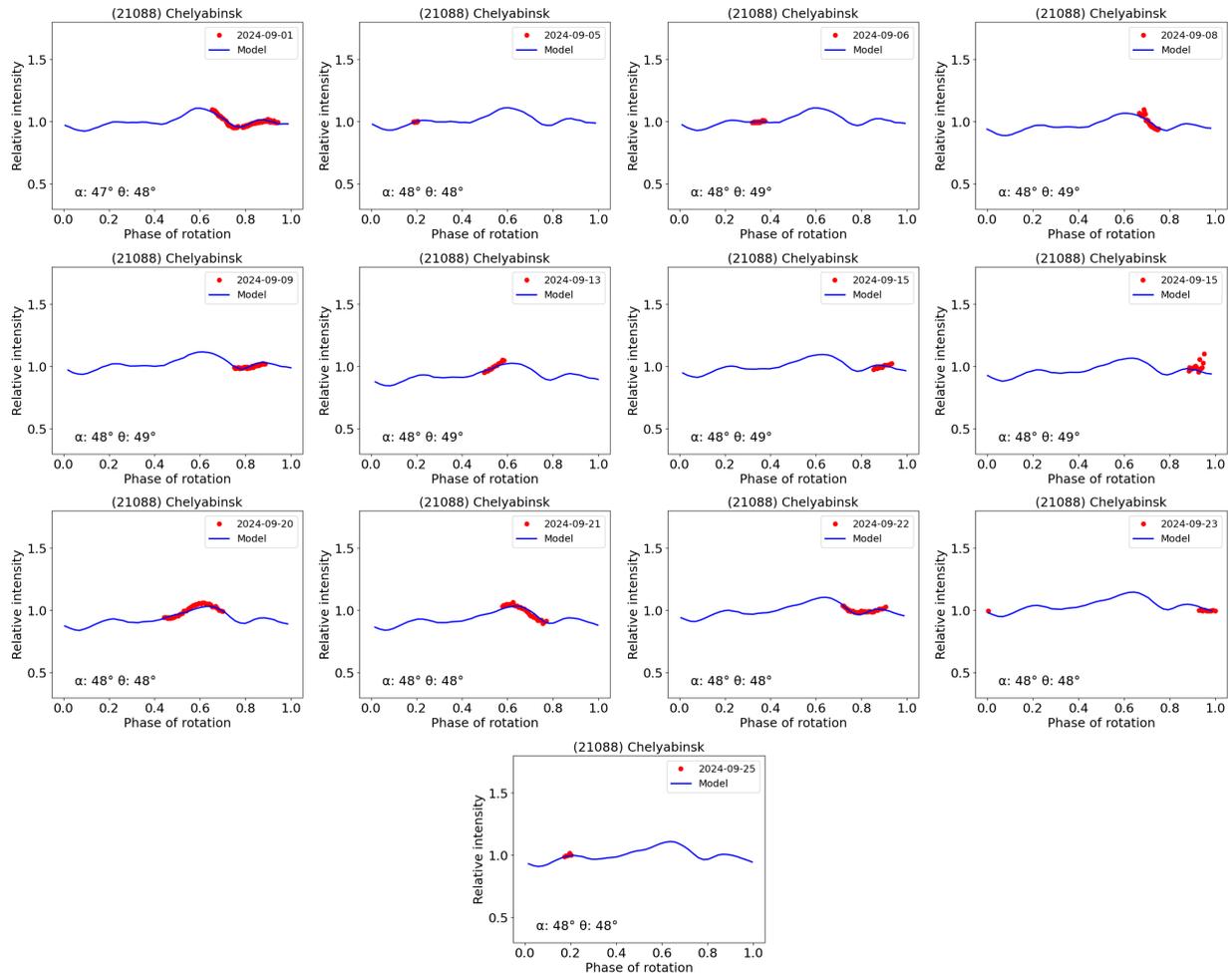

**Figure 25.** (Continued.)

*C.3. (66146) 1998 TU3 Model Fit (Constant Period)*

**Figure 26.** Fit between the rest (35 light curves) of the light curves from (66146) 1998 TU3 used for creating the shape model in this work and the best-fitting constant-period model (C Model). The data are plotted as red circles for each observation, while the model is plotted as a solid blue line. The geometry is described by its solar phase angle $\alpha$ and its aspect angle $\theta$.

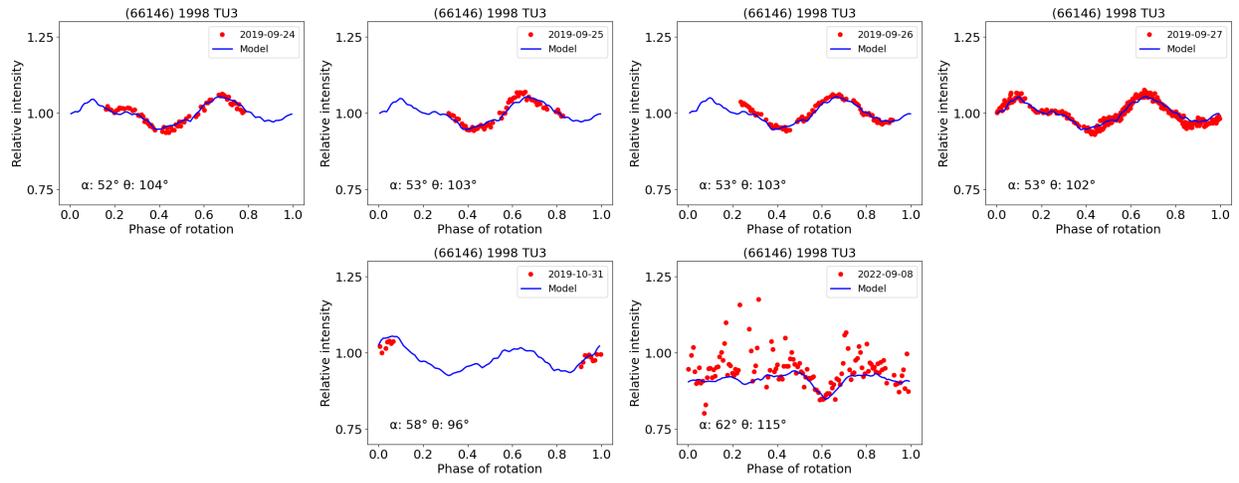

**Figure 26.** (Continued.)

*C.4. (66146) 1998 TU3 Model Fit (Linearly Increasing Period)*

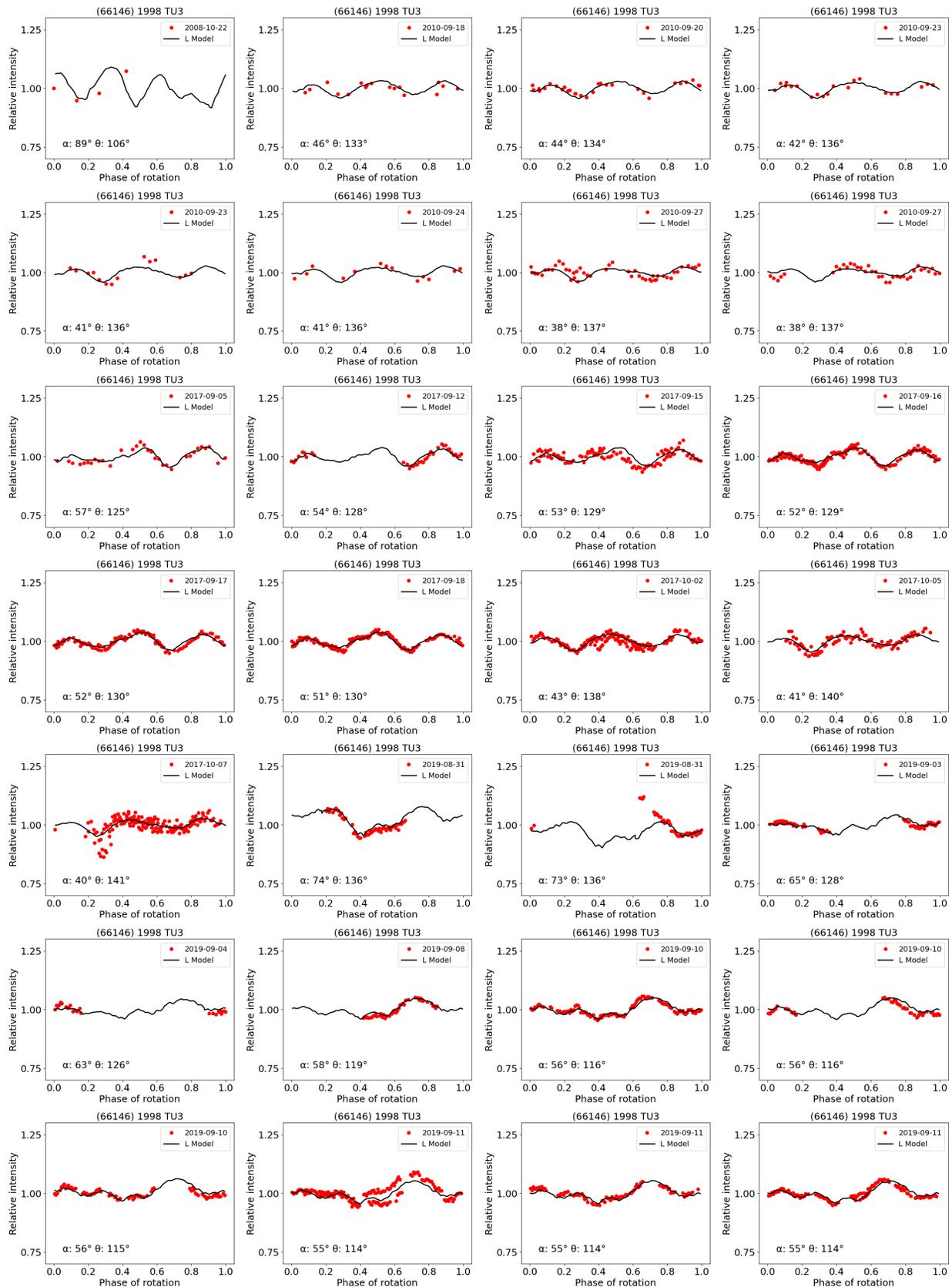

**Figure 27.** Fit between the rest (35 light curves) of the light curves from (66146) 1998 TU3 used for creating the shape model in this work and the best-fitting linearly increasing period model (L Model). The data are plotted as red circles for each observation, while the model is plotted as a solid black line. The geometry is described by its solar phase angle $\alpha$ and its aspect angle $\theta$.

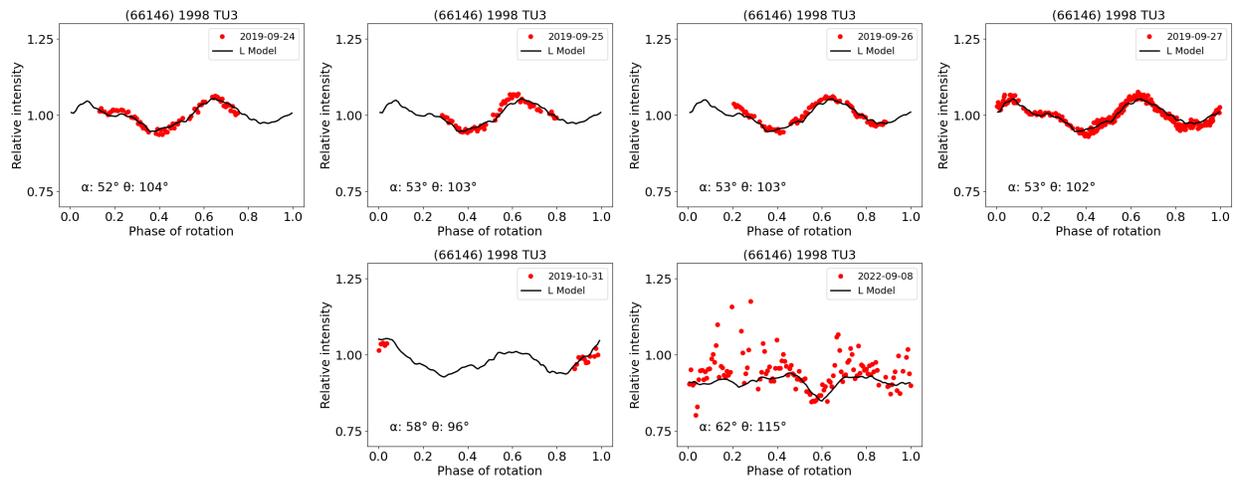

**Figure 27.** (Continued.)

## C.5. (297418) 2000 SP3 Model Fit

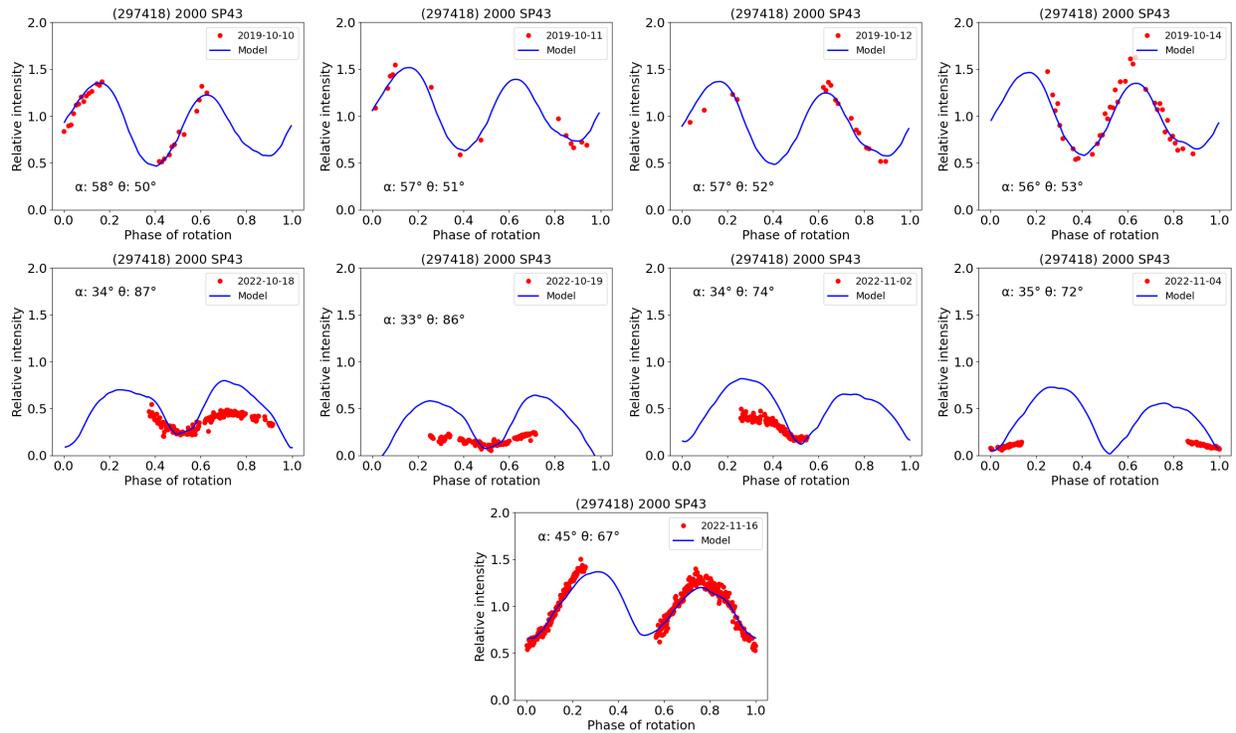

**Figure 28.** Fit between the rest (nine light curves) of the light curves from (297418) 2000 SP43 used for creating the shape model in this work and the best-fitting constant-period model (C Model). The data are plotted as red circles for each observation, while the model is plotted as a solid blue line. The geometry is described by its solar phase angle $\alpha$ and its aspect angle $\theta$.

## ORCID iDs

Javier Rodríguez Rodríguez  https://orcid.org/0000-0002-2718-2022
Enrique Díez Alonso  https://orcid.org/0000-0002-5826-9892
Santiago Iglesias Álvarez  https://orcid.org/0000-0001-5903-9899
Saúl Pérez Fernández  https://orcid.org/0000-0003-4161-778X
Alejandro Buendia Roca  https://orcid.org/0009-0000-9900-4463
Julia Fernández Díaz  https://orcid.org/0009-0007-2419-2405
Javier Licandro  https://orcid.org/0000-0002-9214-337X
Miguel R. Alarcon  https://orcid.org/0000-0002-8134-2592
Miquel Serra-Ricart  https://orcid.org/0000-0002-2394-0711
Amadeo Aznar Macías  https://orcid.org/0000-0002-4351-5157
Francisco Javier de Cos Juez  https://orcid.org/0000-0002-9660-7944